\documentclass[slashbox, 12pt]{article}
\usepackage{amsmath}
\usepackage{amssymb}
\usepackage[singlespacing]{setspace}
\usepackage{natbib}
\usepackage{geometry}
\usepackage{scalefnt}
\usepackage{graphicx}
\usepackage{float}
\usepackage{bibmods}
\usepackage{cite}
\usepackage[notindex,nottoc,notlot,notlof,notbib,chapter,section]{tocbibind}
\usepackage{portland}
\usepackage{lscape}
\usepackage{multirow}
\usepackage{rotating}
\usepackage{slashbox}
\usepackage{epsfig}
\usepackage[titles]{tocloft}
\input{tcilatex}

\def\sepsilon{q_{\varepsilon}}
\def\svarepsilon{\varepsilon}
\def\sgamma{\gamma}
\def\somega{\omega}
\def\sOmega{q_{\omega}}

\def\ssigma{\sigma}

\def\sPi{\Pi}
\def\stheta{\phi} 

\def\se{e}

\def\bbeta{\pmb{\beta}}

\def\bvarepsilon{\pmb{\varepsilon}}
\def\bomega{\pmb{\omega}}

\def\bphi{\pmb{\phi}}

\def\bTheta{\pmb{\Theta}}

\def\be{\pmb{e}}

\def\by{\pmb{y}}

\def\bzero{\pmb{0}}

\def\bI{\pmb{I}}
\def\bX{\pmb{X}}

\def\sgammasq{\sgamma^{2}}
\def\ssigmasq{\ssigma^{2}_{\omega}}
\def\stausq{\ssigma_{\varepsilon}^{2}}

\def\bsigmasq{\pmb{\sigma^{2}_{\omega}}}
\def\btausq{\pmb{\ssigma^{2}_{\varepsilon}}}
\def\bsigsq{\pmb{\sigma^{2}}}

\def\stausqit{\ssigma^{2}_{\varepsilon,i,t}}

\def\stausqstar{\ssigma^{2}_{\varepsilon,\star }}

\def\ssigmasqit{\ssigma^{2}_{\omega,i,t}}

\def\ssigmasqstar{\ssigma^{2}_{\omega,\star }}

\def\ssigsqit{\ssigma^{2}_{i,t}}
\def\ssigsqnit{\ssigma^{2}_{-(i,t)}}
\def\ssigsqint{\ssigma^{2}_{i,-t}}

\def\ssigsqstar{\ssigma^{2}_{\star }}

\def\somegait{\somega_{i,t}}
\def\somegaitm2{\somega_{i,t-2}}
\def\svarepsilonit{\svarepsilon_{i,t}}
\def\svarepsilonitm2{\svarepsilon_{i,t-2}}
\def\somegaik{\somega_{i,k}}
\def\svarepsilonik{\svarepsilon_{i,k}}

\def\svolit{\ssigsqit}

\def\svolitm1{\ssigma^{2}_{i,t-1}}
\def\svolitmq{\ssigma^{2}_{i,t-q}}

\def\sN{ L}
\def\sNi{ L_{i}}
\def\svolt{\ssigma^{2}_{t}}
\def\svoln{\ssigma^{2}_{l}}
\def\svolni{\ssigma^{2}_{l_{i}}}
\def\svolNset{ \{ \ssigma^{2}_{l} \}_{l=1}^{L} }

\def\sPin{\sPi_{l}}
\def\sPini{\sPi_{l_{i}}}

\def\sPiNunboundset{ \{ \sPi_{l} \} }
\def\sPiNiunboundset{ \{ \sPi_{l_{i}} \} }

\def\sshockit{\somega_{i,t},\svarepsilon_{i,t}}

\def\bhomoparams{\bphi,\sgammasq}
\def\bthetaphi{\bphi}
\def\bshockparams{\bomega,\bvarepsilon}
\def\bheteroparams{\bsigsq}
\def\bvolt{\pmb{\svolt}}
\def\bvolnit{\pmb{\ssigsqnit}}
\def\bvolint{\pmb{\ssigsqint}}

\begin{document}
\pagestyle{empty}

\title{Changes in the Distribution of Income Volatility}
\author{Shane T. Jensen and Stephen\ H. Shore\thanks{%
Jensen: Wharton School, Department of Statistics,
stjensen@wharton.upenn.edu. \ Shore: Johns Hopkins University, Department of
Economics, shore@jhu.edu. \ Please contact Shore at: 458 Mergenthaler Hall,
3400 N. Charles Street, Baltimore, MD, 21218; 410-516-5564. } \thanks{%
 }
\thanks{keywords: Markovian hierarchical Dirichlet process, income risk, income volatility, heterogeneity}
\thanks{
We thank Christopher Carroll, Jon Faust, Robert Moffitt, and Dylan Small, for helpful comments,
as well as seminar participants
at the University of Pennsylvania Population Studies Center, the Wharton School,
the 2008 Society of Labor Economists Annual Meeting,
the 2008 Seminar on Bayesian Inference in Econometrics and Statistics,
and the 2008 North American Annual Meeting
of the Econometric Society, and the 2008 Annual Meeting of the Society for Economic Dynamics.} }
\date{}

\maketitle
\thispagestyle{empty}

\begin{abstract}
Recent research has documented a significant rise in the volatility (e.g., expected squared change) of individual incomes
in the U.S. since the 1970s. \ Existing measures of this trend abstract from individual heterogeneity, effectively
estimating an increase in {\em average} volatility. \ We decompose this increase in average volatility and find
that it is far from representative of the experience of most people: there has been no systematic rise in volatility for
the vast majority of individuals. \ The rise in average volatility has been driven almost entirely by a sharp
rise in the income volatility of those expected to have the most volatile incomes, identified {\em ex-ante} by large income
changes in the past. \ We document that the self-employed and those who self-identify as risk-tolerant are much more
likely to have such volatile incomes; these groups have experienced much larger increases in income volatility
than the population at large. \ These results color the policy implications one might draw from the rise in
average volatility. \ While the basic results are apparent from PSID summary statistics, providing a complete
characterization of the dynamics of the volatility distribution is a methodological challenge. \ We resolve these
difficulties with a Markovian hierarchical Dirichlet process that builds on work from the non-parametric Bayesian
statistics literature.

\pagebreak
\end{abstract}

%

\setcounter{page}{1}
\pagestyle{plain}

\doublespace
\section{Introduction}

A large literature argues that income volatility  -- the expectation of squared individual income changes -- has
increased substantially since the 1970s in the U.S., with further increases since the
1990s.\footnote{{\footnotesize  \citet{Dahletal2007} is a noteable exception. \
\citet{Dynanetal2007} provide an excellent survey of research on this subject in their Table 2,
including \citet{GottschalkMoffitt94, GottschalkMoffitt95, DalyDuncan97, DynarskiGruber97, CameronTracy98,
Haider2001, Hyslop2001, GottschalkMoffitt2002, Batchelder2003, Hacker2006, Cominetal2006,
GottschalkMoffitt2006, Hertz2006, Winship2007, BollingerZiliak2007, BaniaLeete2007, Dahletal2007}.
See also \citet{ShinSolon2008}.}}
To the degree that people are risk-averse and income volatility is taken as a proxy for risk,
{\em ceteris paribus} such rising volatility may carry substantial welfare costs. \
As a consequence,
there has been a great deal of recent interest by politicians and journalists in this finding.
\citep{Gosselin2004, Scheiber2004, HouseHearings2007}

To date, research on income volatility trends has ignored individual heterogeneity,
effectively estimating an increase in {\em average} volatility. \ We decompose this increase
in the average and find that it is far from representative of the experience of most people:
there has been no systematic increase in volatility for the vast majority of individuals. \ The increase
has been driven almost entirely by a sharp increase in the income volatility of those with
the most volatile incomes. \
In turn, we find that these individuals with high -- and increasing --
volatility more likely to be self-employed and more likely to self-identify
as risk-tolerant.

Our main finding is apparent in simple summary statistics from the PSID. \ For example, divide the
sample into cohorts, comparing the minority who experienced very large absolute one-year income
changes in the past (e.g., four years ago) to those who did not. \ Since volatility is
persistent, those identified {\em ex-ante} by
large past income changes naturally tend to have more volatile incomes today. \ The income volatility
of this group identified {\em ex-ante} as high-volatility has increased since the 1970s while the income
volatility of others has remained roughly constant.\footnote{{\footnotesize Our finding is consistent
with \citet{Dynanetal2007} who find that increasing income volatility has been driven by the
increasing magnitude of extreme income changes, by the increasingly fat tails of the unconditional distribution of income changes.
\ The fat tails of the unconditional distribution of income changes has also been documented in \citet{GewekeKeane2000}.
\ In its reduced form, our paper shows that
these increasingly fat tails are borne largely by individuals who are \emph{ex-ante}
likely to have volatile incomes.  The increasingly fat tails of the unconditional distribution
are not attributable -- or at least not solely attributable -- to increasingly fat tails of the \emph{expected} distribution
 for everyone.}} \ This divergence of sample moments identifies our key result.

Obviously, these findings could affect substantially the welfare and policy
implications of the rise in average volatility. \ The individuals
whose volatility has increased -- who we find are those with the most volatile incomes -- may be those with the
highest tolerance for risk or the best risk-sharing opportunities. \ Such risk tolerance is apparent not only from
the willingness of these individuals to undertake volatile incomes or self-employment in the first place, but
also from their answers to survey questions.

While the basic results can be seen in summary statistics, providing a complete characterization of the
dynamics of the volatility distribution is a methodological challenge. \ We use a standard model for income
dynamics that allows income to change in response to permanent and transitory shocks. \ What is less standard
is that we allow the variance of these shocks -- our income volatility parameters -- to be heterogeneous and
time-varying.

We estimate a discrete non-parametric model in which volatility parameters are assumed to take one of L
unique values, where the number L and the values themselves are determined by the data. \
We add structure and get tractability with a variant on the Dirichlet process (DP)
prior commonly used in Bayesian statistics. \ The Markovian
hierarchical DP prior model we develop accounts for the grouped nature of the data (by individual) as
well as the time-dependency of successive observations within individuals. \ Implicitly, we place a prior
on the probability that an individual's parameter values will change from one year to the next, on the
number of unique parameter values an individual will hold over his lifetime, and on the number of unique
parameter values found in the sample.

In Section \ref{section: data}, we discuss our data and the summary statistics that drive our results.
\ In Section \ref{section: model}, we present our statistical model including the income process
(Section \ref{section: income process}),
the structure we place on heterogeneity and dynamics in volatility parameters (Section \ref{section: heterogeneity}),
and our estimation strategy (Section \ref{section: estimation}).
\ In Section \ref{section: results}, we show the results obtained by estimating our model on the data. \
Increases in the average volatility parameter are due to increases in volatility among those with
the most volatile incomes
(Section \ref{subsection: pop evol results}). \
We find that the increase in volatility has been greatest among the self-employed and
those who self-identify as risk-tolerant (Section \ref{subsection: whose vol}), and that
these groups are disproportionately
likely to have the most volatile incomes (Section \ref{subsection: who is risky}). \
Increases in risk are present throughout the age distribution, education distribution,
and income distribution (Section \ref{subsection: whose vol}). \
Section \ref{section: conclusion} concludes with a discussion of welfare implications.

\section{Data and summary statistics\label{section: data}}

\subsection{Data and variable construction\label{subsection: data basics}}

Data are drawn from the core sample of the Panel Study of Income Dynamics (PSID).
\ The PSID was designed as a nationally representative panel of U.S. households. \ It
tracked families annually from 1968 to 1997 and in odd-numbered years thereafter;
this paper uses data through 2005. \ The
PSID includes data on education, income, hours worked, employment status,
age, and population weights to capture differential fertility and attrition. \
\ In this paper, we limit the analysis
to\ men age 22 to 60; we use annual labor income as the measure of income.%
\footnote{{\footnotesize Labor income in 1968 is labeled v74 for husbands
and has a constant definition through 1993. \ From 1994, we use the sum of
labor income (HDEARN94 in 1994) and the labor part of business income
(HDBUSY94), with a constant definition through 2005. \ Note that data is
collected on household ``heads" and ``wives" (where the husband is always the
``head" in any couple). \ We use data for male heads so that men who are not household heads (as would be
the case if they lived with their parents) are excluded.}} \ Table \ref{table: sumstat}
presents summary statistics from these data.

{\singlespace
\begin{table}[t]
\caption[Summary Statistics]{Summary Statistics}
\label{table: sumstat}
\begin{center}
{\normalsize
\begin{tabular}{l c c c c}
\hline \hline
& mean & st. dev. & min & max \\ \cline{2-5}
year & $1986.3$ & $10.0$ & $1968$ & $2005$ \\ 
age (years) & $~40.0~$ & $~10.5~$ & $22$ & $60$ \\ 
education (years) & $~13.1~$ & $~2.9~$ & $0$ & $17$ \\ 
\# of observations/person & $17.2$ & $9.0$ & $1$ & $34$ \\ 
married (1 if yes, 0 if no) & $~0.80~$ & $.$ & $.$ & $.$ \\ 
black (1 if yes, 0 if no) & $0.05$ & $.$ & $.$ & $.$ \\ 
annual income (2005 \$s) & $\$50,553$ & $\$57,506$ & $0$ & $\$3,714,946$ \\ 
annual income (\$s) & $\$29,277$ & $\$46,818$ & $0$ & $\$3,500,000$ \\ 
family size & $3.1$ & $1.5$ & $1$ & $14$ \\ \hline
\end{tabular}
}
\end{center}
\par
{\footnotesize This table summarizes data from 52,181 observations on 3,041
male household heads.  }
\end{table}
}

We want to ensure that changes in income are not driven by changes in the
top-code (the maximum value for income entered that can be entered in the
PSID). \ The lowest top code for income was \$99,999 in 1982
(\$202,281 in 2005 dollars), after which the top-code rises to
\$9,999,999. \ So that top-codes will be standardized in real terms, this
minimum top-code is imposed on all years in real terms, so the top-code is
\$99,999 in 1982 and \$202,281 in 2005. \ Since our income process in Section \ref{section: income process}
does not model unemployment explicitly, we need to ensure that
results for the log of income are not dominated by small changes in the
level of income near zero (which will imply huge or infinite changes in the
log of income). \ To address this concern, we replace income values that are
very small or zero with a non-trivial lower bound. \ We choose as this
lower-bound the income that would be earned from a half-time job (1,000
hours per year) at the real equivalent of the 2005 federal minimum wage
(\$5.15 per hour). \ This imposes a bottom-code of \$5,150 in 2005 and
\$2,546 in 1982. \ Note that the difference in log income between the top-
and bottom-code is constant over time, so that differences over time
in the prevalence of predictably extreme income changes cannot be driven by changes in the possible range of
income changes. \ The vast majority of the values
below this bound are exactly zero. \ This bound allows us to exploit
transitions into and out of the labor force. \ At the same time, the bound
prevents economically unimportant changes that are small in levels
but large and negative in logs from
dominating the results. \ Results are robust to other values for this lower
bound, such as the income from full-time work (2,000 hours per year) at the
2005 minimum wage (in real terms).\footnote{%
The Winsorizing strategy employed here is obviously second-best to a strategy of
modeling a zero income explicitly. \ Unfortunately, such a model is not feasible
given the complexity added by evolving and heterogeneous volatility parameters. \
The other alternative would be simply to drop observations with low incomes, though we view this approach is
much more problematic in our context; it would explicitly rule out the extreme income changes that are the subject of this paper.}

{\singlespace
\begin{table}[t]
\caption[Distribution of Income]{Distribution of Income, Excess Log Income,
and Income Changes for Men}
\label{table: ex distribution}
\begin{center}
\begin{tabular}{l c c c c c}
\hline \hline

& Real Income &&  \multicolumn{3}{c}{Excess Income} \\ \cline{2-2} \cline{4-6}
 & Level  &&  Level  & One-Year & Five-Year \\  
$\begin{array}{c} \rm{One-Year} \\ \rm{Change} \end{array}$ &
$\begin{array}{c} \rm{Five-Year} \\ \rm{Change} \end{array}$ \\  \hline
Mean                            & \$50,553 (\$48,867) && 0 & 0.0017 & 0.0043           \\ 
St. Dev.                        & \$57,506 (\$34,943) && 0.7307 & 0.4870 & 0.6863      \\ 
Observations                    & 52,181  && 52,181 & 43,261 & 34,972                  \\ 
Minimum                         & \$0 (\$5,150) && -2.9325 & -3.6877 & -3.8361         \\ 
5$^{\rm{th}}$ Percentile      & \$668  (\$5,150) && -1.6283 & -0.7323 & -1.3046     \\ 
25$^{\rm{th}}$ Percentile     & \$26,174 && -0.2964 & -0.1089 & -0.2126              \\ 
50$^{\rm{th}}$ Percentile     & \$42,887  && 0.1246 & 0.0134 & 0.0653                \\ 
75$^{\rm{th}}$ Percentile     & \$62,012  && 0.4601 & 0.1442 & 0.3072                \\ 
95$^{\rm{th}}$ Percentile     & \$113,500  && 0.9757 & 0.6673 & 0.9764               \\ 
Maximum                         & \$3,714,946 (\$202,381) && 2.6435 & 3.5862 & 4.0678  \\ \hline
\end{tabular}
\end{center}
\par
{\footnotesize Table \ref{section: data} describes the distribution of labor
income for men in the PSID over the period from 1968 to 2005. \ See Section %
\ref{section: data} for a detailed description of the income variable and
the top- and bottom-coding procedure. Column 1 shows the distribution of
real annual income for men (in 2005 dollars). \ The numbers in parentheses are the values with
top- and bottom-coding restrictions. \ Column 2 shows the distribution of
\textquotedblleft excess\textquotedblright\ log income, the residual from
the regression of log labor income (with top- and bottom-code adjustments)
 on the covariates enumerated in Section %
\ref{section: data}. \ Column 3 presents the distribution of one-year
changes in excess log income. \ Column 4
repeats the results for column 3, but presents five-year changes instead of
one-year changes.  }
\end{table}
}

In this paper, we model the evolution of \textquotedblleft
excess\textquotedblright\ log income. \ This is taken as the residual from a
regression to predict the natural log of labor income (top- and bottom-coded as
described). \ The regression is weighted by the PSID-provided sample weights,
with the weights normalized so that the average weight in each year is the same. \
We use as regressors: a cubic in age for
each level of educational attainment (none, elementary, junior high, some
high school, high school, some college, college, graduate school); the
presence and number of infants, young children, and older children in the
household; the total number of family members in the household, and dummy
variables for each calendar year. \ Including calendar year dummy variables
eliminates the need to convert nominal income to real income explicitly. \
While this step is standard in the income process
literature, it is not necessary to obtain our results. \ The results to follow are
qualitatively the same and quantitatively similar when we use log income in lieu
of excess log income.

Table \ref{table: ex distribution} presents data on the distribution of real
annual income in column 1 (imposing top- and bottom-code
restrictions in parentheses). While the mean real income is nearly identical
with and without top- and bottom-code restrictions (\$50,553 versus
\$48,867), these restrictions on extreme values reduce the standard
deviation of real income from \$57,506 to \$34,943. \ Column 2 shows the
distribution of \textquotedblleft excess\textquotedblright\ log income. \
Since excess log income is the residual from a regression, its mean is zero.
\ The inter-quartile range of excess log income is $-0.30$ to $0.46$. \

Column 3 presents the distribution of one-year changes in
excess log income. \ Naturally, the mean of one-year changes is close to
zero. \ The inter-quartile range of one-year changes is $-0.11$ to $0.14$;
excess income does not change more than $11$ to $14$ percent from year to
year for most individuals. \ However, there are extreme changes in income,
so the standard deviation of changes to log income ($0.49$) is far great than the
inter-quartile range. \ This implies
either that changes to income have fat tails (so that everyone faces
a small probability of an extreme income change), or alternatively that there is
heterogeneity in volatility (so that a few people face a non-trivial probability
of an extreme income change). \ Unless a model is identified from parametric
assumptions, these are observationally equivalent in a
cross-section of income changes. \ However, heterogeneity and fat tails
have different implications for the time-series of volatility, and we
exploit these in the paper.

Column 4 repeats the results from column 3, but presents five-year excess log income changes instead of one-year changes. \
These long-term changes have only slightly higher standard deviations than the one-year change, $0.69$ vs. $0.49$,
suggesting some mean-reversion in income. \  \citet{AbowdCard89} show that while one-year income changes are highly
negatively correlated at one-year lags, there is no evidence of autocorrelated income changes at lags greater than two years.

\subsection{Volatility summary statistics\label{subsection: moments basics}}

{\singlespace
\begin{table}[p]
\caption[Income Volatility Sample Moments]{Income Volatility Sample Moments}
\label{table: momentsyby}\begin{center}
{\small
\begin{tabular}{ l c c c c c c c}
\hline \hline
\setlength{\tabcolsep}{1pt}
 & \multicolumn{3}{c}{Permanent Variance} &&  \multicolumn{3}{c}{Squared Change} \\ \cline{2-4} \cline{6-8}
  &       Mean &       Median &       95$^{\rm{th }}\%$ &&       Mean        &       Median &       95$^{\rm{th }}\%$ \\  \hline
Average &       0.1091  &       0.0099  &       0.8264  &&       0.3561  &       0.0314  &       2.0042  \\      
\% Change  &      \multirow{2}{*}{49$\%$}      &    \multirow{2}{*}{15$\%$}    &     \multirow{2}{*}{92$\%$}      &&
           \multirow{2}{*}{110$\%$}       &     \multirow{2}{*}{19$\%$}    &   \multirow{2}{*}{143$\%$ } \\
1970-2003 &           &              &             &&            &             &           \\      
Slope   &       0.0015  &       0.0000  &       0.0205  &&       0.0106  &       0.0002  &       0.0775   \\
(t-stat)        &        (4.11) &        (0.52) &        (8.76) &&        (11.96)&       (1.26)  &       (11.18)    \\      \hline  
1970    &   .       &       .       &       .       &&      0.1555  &       0.0210  &       0.7709     \\      
1971    &   .       &       .       &       .       &&      0.1823  &       0.0229  &       0.8004     \\      
1972    &   0.0665  &       0.0059  &       0.4003  &&      0.2142  &       0.0277  &       1.1276     \\      
1973    &   0.0786  &       0.0048  &       0.4423  &&      0.2296  &       0.0269  &       1.1500     \\      
1974    &   0.0792  &       0.0054  &       0.5090  &&      0.2324  &       0.0264  &       1.1059     \\      
1975    &   0.0986  &       0.0129  &       0.6243  &&      0.2496  &       0.0380  &       1.2286     \\      
1976    &   0.0997  &       0.0179  &       0.6749  &&      0.3124  &       0.0498  &       1.6006     \\      
1977    &   0.0933  &       0.0095  &       0.7058  &&      0.2983  &       0.0316  &       1.8058     \\      
1978    &   0.0706  &       0.0062  &       0.5958  &&      0.2751  &       0.0296  &       1.3344     \\      
1979    &   0.0838  &       0.0061  &       0.6415  &&      0.2931  &       0.0269  &       1.6711     \\      
1980    &   0.1388  &       0.0115  &       0.9270  &&      0.2811  &       0.0292  &       1.4495     \\      
1981    &   0.1159  &       0.0123  &       0.8844  &&      0.2932  &       0.0296  &       1.5200     \\      
1982    &   0.1004  &       0.0150  &       0.7256  &&      0.2514  &       0.0305  &       1.2840     \\      
1983    &   0.0859  &       0.0150  &       0.6630  &&      0.2912  &       0.0330  &       1.5820     \\      
1984    &   0.1220  &       0.0126  &       0.8786  &&      0.3185  &       0.0331  &       1.8609     \\      
1985    &   0.1109  &       0.0118  &       0.7869  &&      0.3283  &       0.0370  &       1.7499     \\      
1986    &   0.1002  &       0.0110  &       0.6905  &&      0.3089  &       0.0358  &       1.5483     \\      
1987    &   0.1089  &       0.0093  &       0.7739  &&      0.3015  &       0.0295  &       1.6058     \\      
1988    &   0.1224  &       0.0087  &       0.7969  &&      0.3121  &       0.0300  &       1.6476     \\      
1989    &   0.1161  &       0.0077  &       0.8171  &&      0.3278  &       0.0276  &       1.8996     \\      
1990    &   0.1174  &       0.0091  &       0.7770  &&      0.2998  &       0.0261  &       1.5937     \\      
1991    &   0.1312  &       0.0121  &       0.9905  &&      0.3523  &       0.0309  &       1.8485     \\      
1992    &   0.1013  &       0.0111  &       0.9119  &&      0.3168  &       0.0295  &       1.7572     \\      
1993    &   0.1272  &       0.0112  &       1.0935  &&      0.4166  &       0.0333  &       2.3561     \\      
1994    &   0.1083  &       0.0104  &       0.9270  &&      0.4479  &       0.0347  &       2.6530     \\      
1995    &   0.1346  &       0.0077  &       1.1290  &&      0.4914  &       0.0333  &       3.3055     \\      
1996    &   .       &       .       &       .       &&      0.4768  &       0.0264  &       3.1923     \\      
1997    &   0.0898  &       0.0074  &       0.8660  &&      0.4671  &       0.0282  &       2.9644     \\      
1999    &   0.1142  &       0.0080  &       0.9632  &&      0.4539  &       0.0317  &       2.7189     \\      
2001    &   0.1190  &       0.0073  &       1.1174  &&      0.4463  &       0.0271  &       2.9567     \\      
2003    &   0.1487  &       0.0182  &       1.2951  &&      0.6348  &       0.0574  &       3.9098     \\      \hline

\end{tabular}
}
\end{center}
\par
{{\footnotesize The year $t$ permanent variance is the product of two-year changes in excess log income (from $t-2$ to $t$)
and the six-year changes that span them
(from $t-4$ to $t+2$). \ The year $t$ squared change is from $t-2$ to $t$. \
The first row shows full sample moments. \ The second row shows the
percent change over the sample, calculated as the coefficient of
a weighted OLS regression of year-specific sample moments on a time trend, multiplied by the number of years (2005-1968) and
divided by the full sample moment. \ The coefficient and t-statistic are shown below.  } }
\end{table}
}

Table \ref{table: momentsyby} shows the evolution of
volatility sample
moments over time. \ The first
three columns show the variance of permanent income changes.\footnote{\footnotesize{The
variance of permanent income changes
is the individual-specific product of two-year changes in excess log income
(for example, between years $t$ and $t-2$) and the six-year changes that span them (for example, between years $t+2$
and $t-4$).
\ \citet{MeghirPistaferri2004} show that this moment identifies the variance of permanent income changes (between years t-2 and t)
under fairly general
conditions, including the income process we use in Section \ref{section: income process}.}}
\ The final three columns
present two-year squared changes in excess log income, a raw measure of income
volatility.\footnote{\footnotesize{All use weights from
the PSID. \ The first row shows whole-sample results. \ The second row shows the
percent change in the mean, median, or 95$^{\rm{th}}$ percentile over the sample. \ This is merely calculated as coefficient of
a weighted OLS regression of the year-specific sample moment on a time trend, multiplied by the number of years ($2005-1968$) and
divided by the whole-sample value in the previous row. \ The coefficient and t-statistic from this regression are shown just below. \
Year-by-year values are then shown.}}
Note that while the mean size of an income change (columns 1 and 4, Table \ref{table: momentsyby})
has increased over time, the median (columns 2 and 5) has not. \ This divergence can
be explained by an increase in the magnitude of large unlikely income changes (columns 3 and 6).
\ While not framed
in this way, these features of the data have been identified in previous research, including \citet{Dynanetal2007}. \

{\singlespace
\begin{table}[p]
\caption[Income Volatility Sample Moments by Past Volatility]{Income Volatility Sample Moments by Past Volatility}
\label{table: persistmomyby}\begin{center}
{\small
\begin{tabular}{lccccc}
\hline\hline
 & \multicolumn{2}{c}{Permanent Variance} &&  \multicolumn{2}{c}{Squared Change} \\ \cline{2-3}\cline{5-6}
Moment  &       \multicolumn{2}{c}{Mean} &&       \multicolumn{2}{c}{Mean} \\      
Past Variance &       Low &       High&&       Low &       High \\      \hline
Average &       0.0820  &       0.3845  &&       0.2675  &       0.6879  \\      
$\%$ Difference &\multicolumn{2}{c}{92$\%$}&&\multicolumn{2}{c}{54$\%$}      \\      
Slope   &       0.00083 &       0.020   &&       0.0080  &       0.026   \\
(t-stat)&       (1.29)  &       (4.36)  &&       (8.67)  &       (6.61)   \\      \hline  
1974    &       .       &       .       &&       .       &       .       \\      
1975    &       .       &       .       &&       .       &       .       \\      
1976    &       0.1015  &       0.2895  &&       0.2265  &       0.5304  \\      
1977    &       0.0935  &       0.3260  &&       0.2164  &       0.5917  \\      
1978    &       0.0374  &       0.1955  &&       0.1540  &       0.3231  \\      
1979    &       0.0491  &       0.3720  &&       0.2017  &       0.4381  \\      
1980    &       0.0786  &       0.3663  &&       0.1972  &       0.5860  \\      
1981    &       0.0668  &       0.2558  &&       0.1780  &       0.5981  \\      
1982    &       0.0608  &       0.2214  &&       0.1964  &       0.5569  \\      
1983    &       0.0676  &       0.0927  &&       0.1806  &       0.5065  \\      
1984    &       0.1285  &       0.3449  &&       0.2426  &       0.4804  \\      
1985    &       0.0757  &       0.2262  &&       0.2708  &       0.4550  \\      
1986    &       0.1178  &       0.0190  &&       0.2210  &       0.6276  \\      
1987    &       0.0753  &       0.3392  &&       0.2532  &       0.4401  \\      
1988    &       0.0600  &       0.2691  &&       0.2381  &       0.6474  \\      
1989    &       0.0701  &       0.3087  &&       0.2430  &       0.6448  \\      
1990    &       0.0964  &       0.4907  &&       0.2143  &       0.3365  \\      
1991    &       0.1108  &       0.4253  &&       0.2846  &       0.8574  \\      
1992    &       0.0783  &       0.3356  &&       0.2498  &       0.5450  \\      
1993    &       0.0889  &       0.6556  &&       0.2990  &       0.8766  \\      
1994    &       0.0569  &       0.2607  &&       0.3339  &       0.7283  \\      
1995    &       0.1105  &       0.5464  &&       0.3327  &       0.8622  \\      
1996    &       .       &       .       &&       0.3590  &       0.8988  \\      
1997    &       0.0428  &       0.9663  &&       0.3375  &       0.8572  \\      
1999    &       0.0865  &       0.6554  &&       0.3309  &       1.2439  \\      
2001    &       0.1049  &       0.4295  &&       0.3115  &       1.0118  \\      
2003    &       0.1101  &       0.8355  &&       0.4460  &       1.2074  \\      \hline
\end{tabular}
}
\end{center}
\par
{{\footnotesize The year $t$ permanent variance is the product of two-year changes in excess log income (from $t-2$ to $t$)
and the six-year changes that span them
(from $t-4$ to $t+2$). \ The first and third columns show sample means for the cohort of individuals whose permanent variance and
squared change, respectively,
were below median in the year four years prior. \ The second and fourth columns show the same, but for the cohorts
with past values above the 95$^{\rm{th}}$ percentile four years prior. \
The first row shows full sample moments. \ The third and fourth rows present the coefficient and t-statistic
from a weighted OLS regression of year-specific sample means on a time trend. \ The difference in these two coefficients,
divided by their average, is the $\%$ difference in the second row. \ Year-by-year means are shown below.  } }
\end{table}
}
{\singlespace
\begin{figure}[t]
\caption[Evolution of Persistent Volatility]{Comparing Sample Variances for Those
With and Without Large Past Income Changes}
\label{fig:volpersist}
\begin{center}
{\normalsize
\begin{tabular}{cc}
Permanent Variance  & Squared Change \\
\includegraphics[width=2.8in]{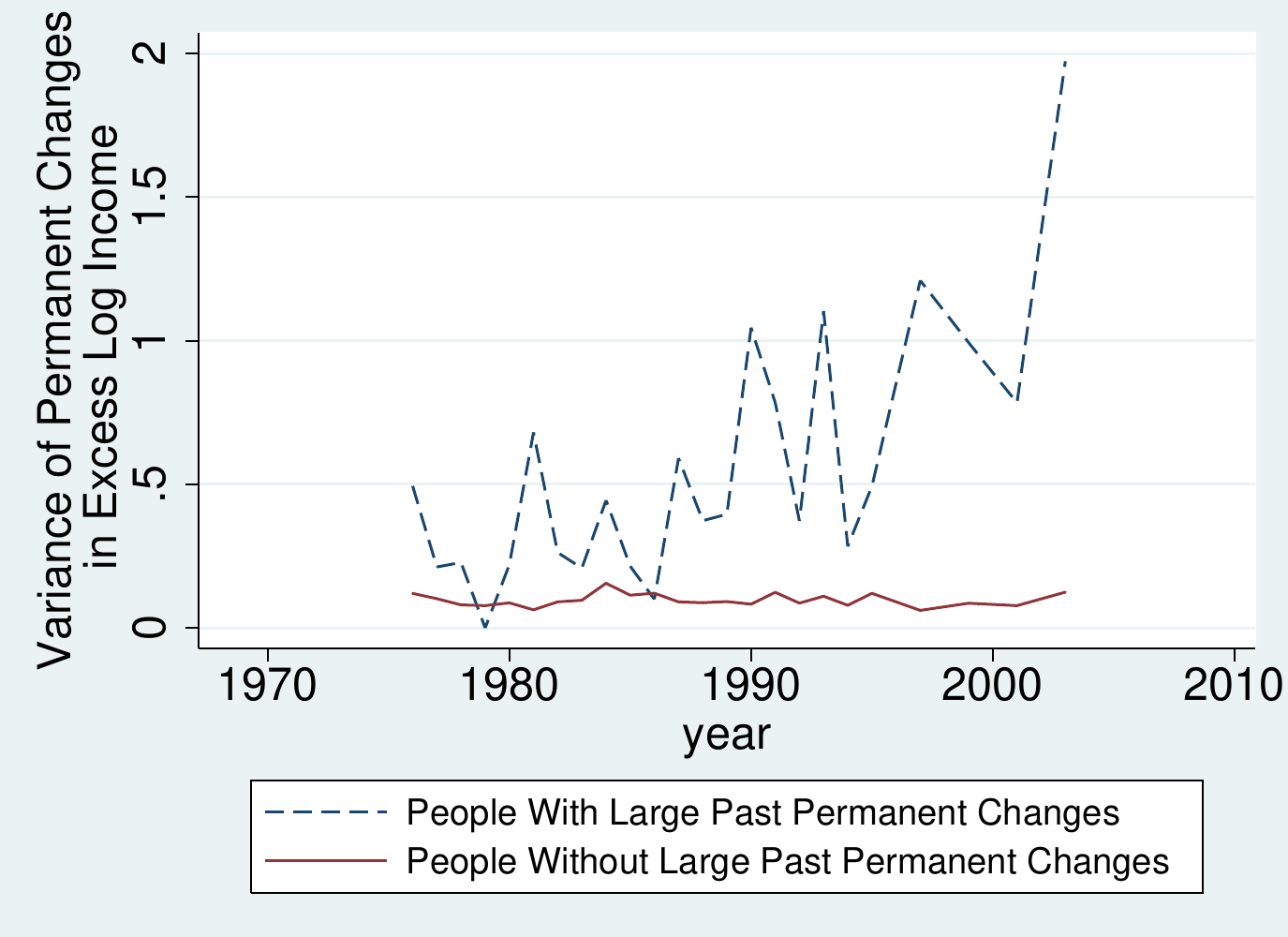} &
\includegraphics[width=2.8in]{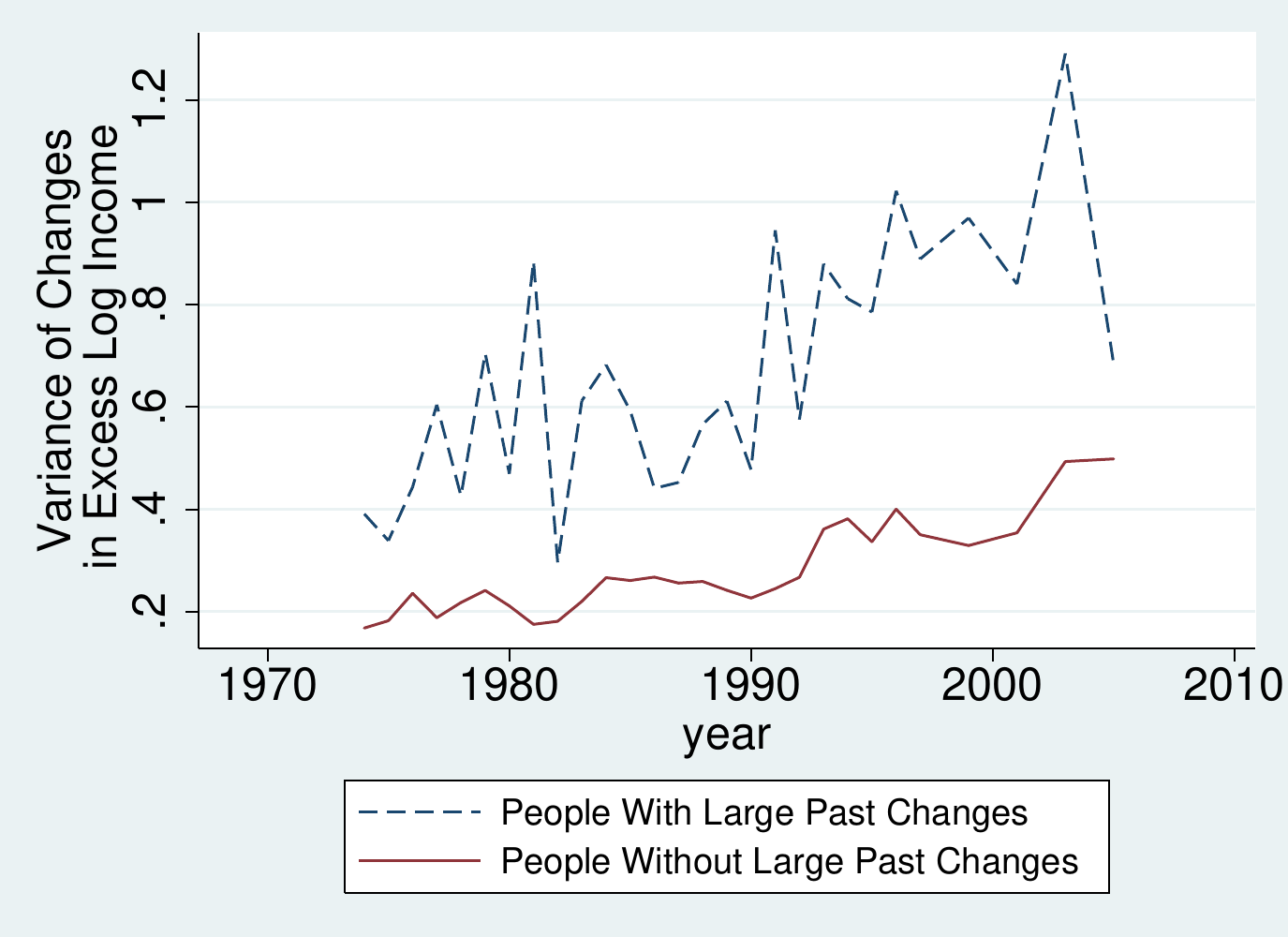}  \\
\end{tabular}
}
\end{center}
\par
{\footnotesize Following \citet{MeghirPistaferri2004}, the sample permanent variance is calculated as the product of two-year changes in
excess log incomes (between years t and t-2) and the six-year changes that span them
(between years t+2 and t-4). \ The sample transitory variance is calculated as the square of two-year
changes in excess log income. \ Individuals are defined as low past variances when their sample
variance (permanent or transitory, respectively) four years ago is below median; individuals are defined as high past variance when
their sample variance four years ago is above the 95$^{\rm{th}}$ percentile. \ Weighted averages for these groups are presented
in each year for which data is available for permanent variance (left panel) and transitory variance (right panel). \\ \line(1,0){430}}
\end{figure}
}

Table \ref{table: persistmomyby} and Figure \ref{fig:volpersist} show
the evolution of volatility sample moments separately for those
who are {\em ex-ante} likely or unlikely to have volatile incomes. \
The left panel of Table \ref{table: persistmomyby} presents the sample
mean of the permanent variance; the right panel
presents the mean two-year squared excess log income change. \
For each year, the sample is split into two groups (below median or above 95$^{\rm{th}}$ percentile)
based on the absolute magnitude of permanent (left panel)
or squared (right panel) changes four years prior. \
Unsurprisingly, individuals with large past income changes tend to have larger subsequent income changes. \
The tendency to have large income changes is persistent, which indicates that some
individuals have \emph{ex-ante} more volatile incomes than others.

If (as we argue) volatility is increasing for high-volatility individuals but not for low-volatility
individuals, then the gap in the sample variance between those with and without large past income
changes should be increasing over time. \ This divergence over time in volatility between
past low- and high-volatility cohorts is clear in both Table \ref{table: persistmomyby} and Figure \ref{fig:volpersist}. \
The magnitude of income changes has been increasing more for those
with large past income changes (who are more likely to be inherently high-volatility) than for those
without such large past income changes (who are not).  \ This is particularly apparent for the permanent variance; for the transitory variance,
the finding is obscured slightly by the jump in volatility for everyone in the early- to
mid-nineties (when the PSID changed to an automated data collection system which may have led to increased
measurement error in income). \ This divergence illustrates the key stylized fact developed in this paper: the increase in
income volatility can be attributed to an increase in volatility among those with the most volatile incomes,
identified {\emph ex-ante} by large past income changes.

\section{Statistical model\label{section: model}}
\subsection{Income process\label{section: income process}}

Here, we present a standard process for excess log income for individual $i$ at time $t$
\citep[following][and
many others]{CarrollSamwick97,MeghirPistaferri2004}:

\begin{eqnarray}
y_{i,t} &=&p_{i,t}+\xi_{i,t}+\se_{i,t}  \label{eq: income process} \\
p_{i,t} &=&p_{i,0}+\sum\limits_{k=1}^{t-\sOmega }\somegaik
+\sum\limits_{k=t-\sOmega +1}^{t}\stheta _{\omega,t-k}\somegaik\rm{.}
\notag  \\
\xi _{i,t} &=&\sum\limits_{k=t-\sepsilon +1}^{t}\stheta _{\varepsilon,t-k}\svarepsilonik
\notag
\end{eqnarray}%
Excess log income ($y_{i,t}$) is the sum of permanent income ($p_{i,t}$), transitory
income ($\xi _{i,t}$), and measurement error ($\se_{i,t}$). \
\ The permanent shock, transitory shock, and measurement error are
assumed to be normally distributed with mean zero as well as
independent of one another, over time and across individuals.
\ Permanent
income is initial income ($p_{i,0}$) plus the weighted sum of past permanent shocks
($\somega_{i,k}, 0 < k \le t$) with variance $\ssigmasqit\equiv E\left[
\somegait^{2}\right] $. \
Transitory income is the weighted sum of recent
transitory shocks ($\svarepsilon_{i,k}$) with variance $\stausqit\equiv E\left[ \svarepsilonit^{2}\right] $.
 \ We refer to $\ssigsqit \equiv (\stausqit,\ssigmasqit)$
jointly as the volatility parameters. \ These will be allowed to differ between individuals to accommodate heterogeneity,
and to evolve over time. \ This accommodates not just an evolving distribution of volatility parameters, but also
systematic changes over the life-cycle in volatility paramters, as suggested by \citet{ShinSolon2008}. \
Subcripts for $i$ and $t$ indicate that volatility parameters
may differ across individuals and over time, as discussed in Section \ref{section: heterogeneity}. \
\textquotedblleft Noise variance\textquotedblright refers to the
variance of measurement error, $\sgammasq\equiv E\left[
\se_{i,t}^{2}\right] $. \
This measurement error could be subsumed into
transitory income; it is kept separate only to accommodate our estimation strategy.

Here, permanent shocks come into effect over $\sOmega $
periods, and transitory shocks fade completely after $\sepsilon$ periods.\footnote{%
In \citet{CarrollSamwick97},  $\stheta_{\omega,k}=\stheta_{\varepsilon,k}=0$ is assumed
for $k>0$, though the authors acknowledge that this assumption is unrealistic
and design an estimation strategy that is robust to this restriction but do not estimate $\stheta_{k}$.
\ In \citet{MeghirPistaferri2004} and \citet{Blundelletal2008}, $\stheta_{\omega,k}=0$ is assumed
for $k>0$ but $\stheta_{\varepsilon,k}=0$ is not.}
\ As an example of our notation, $\stheta_{\omega,2}$ denotes the weight placed on a permanent shock
from two periods ago, $\somegaitm2$,  in current excess log income;
$\stheta_{\varepsilon,2}$ denotes the weight placed on a transitory shock from two periods ago,
$\svarepsilonitm2$,  in current excess
log income. \
While we use the word \textquotedblleft shock\textquotedblright\ for
parsimony, these innovations to income may be predictable to the individual, even
if they look like shocks in the data. \ Without loss of generality, we impose the constraint
that the weights placed on transitory shocks sum to one ($\sum_{k}\stheta_{\varepsilon,k}=1$).

\subsection{Heterogeneity and dynamics\label{section: heterogeneity}}

We characterize the dynamics of volatility parameters, $\svolit$, using a discrete non-parametric approach. \
In a discrete non-parametric model, the variable of interest -- here, the pair $\svolit \equiv (\stausqit,\ssigmasqit) $
-- can take one of $\sN$ possible values, $\svolNset$ (where $\sN$ and $\svolNset$ for any given sample are determined by the data). \
The probability that $\svolit$ takes a given value is a function of a) the distribution of values in the population,
$\sPiNunboundset$, where $\sPin$ is the proportion of the population whose parameter values are equal to $\svoln$, b) the distribution of values for each individual $i$, $\sPiNiunboundset$, where $\sPini$ is the proportion of individual $i$'s observations with parameter values are equal to $\svoln$,, and c) the number of consecutive years $Q_{i,t}$ with the most recent value.\footnote{\footnotesize{$Q_{i,t}$ is the largest value satisfying $\svolitm1=\svolitmq$ for all $0 < q_{it} \le Q_{i,t}$.}} \ In other words, $\svolit$ has a given
probability of changing from one year to the next; when it changes, it changes to a value drawn from the individual's
distribution,
 $\sPiNiunboundset$,
 which in turn consists of values drawn from the population distribution, $\sPiNunboundset$.

We add structure and get tractability
by adding a prior commonly used in Bayesian analysis of such discrete non-parametric problems:
the Dirichlet process (DP) prior. \ In a standard DP model, there is a ``tuning parameter'', $\Theta$, which implicitly places a prior on the total number of unique parameter values in the sample, $\sN$.\footnote{\footnotesize{In large samples the expected number of unique values is of the order $\Theta \log((N + \Theta)/\Theta)$ where $N$ is the number of observations. \citep{Liu96}}} \ $\Theta$ is defined more formally in Section \ref{section: estimation}. \ We set $\Theta = 1$, though our inference is not sensitive to this choice. \ In a hierarchical DP (HDP) model \citep[recently developed by][]{TehJorBea06}, the usual DP model is extended so by adding a second tuning parameter, $\Theta_{i}$,  which implicitly places a prior on the total number of unique parameter values for any given individual, $\sNi$; we set $\Theta_{i}=1$.

We extend this approach further to address panel data by including a Markovian structure on the hierarchical DP, giving us a Markovian hierarchical DP (MHDP) model. \ In our Markovian approach, the prior probability that the parameter is unchanged from the previous period depends on the number of consecutive years with that value, $Q_{i,t}$. \ We add a third tuning parameter, $\theta$, to place a prior on the probability of changing the parameter value, $p \left( \svolit =\svolitm1 \ | i,t \right)=Q_{i,t} / (\theta+ Q_{i,t})$; we set $\theta=1$. \ In the MHDP model, our prior parameters can then be characterized with the triple $\bTheta\equiv\{\Theta,\Theta_{i},\theta\}=\{1,1,1\}$.

Given our research question, a key advantage of this set-up is that it does not restrict the shape (or the evolution of the shape) of the cross-sectional volatility distribution. \ We view our discrete non-parametric model and the structure placed on it by our MHDP prior as providing a sensible middle ground between tractability and flexibility.

\subsection{Estimation\label{section: estimation}}

We estimate the income process from Section \ref{section: income process}
on annual data from the PSID (detailed in Section \ref{section: data})
for excess log income. \
When data are missing, mostly because no data was collected by the PSID
in even-numbered years following 1997, we impute bootstrapped guesses of income.\footnote{\footnotesize{We examine the
two-year change in excess log income that spans any single-year of missing data. \ We identify the set of two-year
excess log income changes with a similar magnitude elsewhere in the data and select one at
random. \ This bootstrapped
draw has an intermediate value which is used to fill in the missing data. \ For example, consider
an individual with excess log income of 0.1 in 1999, 0.5 in 2001 and
(since the PSID did not gather data in the intervening year)
missing in 2000. \ From the set of all sample observations with two-year excess
log income changes in the neighborhood of 0.4,
we select one at random. \ In general, this observation will be drawn from a different individual
than the one with the missing data. \ Imagine that the individual-years drawn at random
have excess log incomes of 0.6, 0.7, and 1.0 in 1972, 1973, and 1974,
respectively. \ We then fill in the original individual's missing data in 2000 with 0.2 (0.1+0.7-0.6). \
We drop individuals with longer spans of missing data.}} \ These bootstrapped values
add no additional information; they merely accommodate our estimation strategy in a setting with missing data in a way
that is intended to minimize the possible impact on our results. \
Here, we outline an approach for combining the prior from Section \ref{section: heterogeneity} with
data on excess log income, $\by$, to form a posterior on the distribution of volatility parameters,
$\bheteroparams$.\footnote{\footnotesize{$\by$ is the ragged $N$ by $T+1$ matrix, with $y_{i,t}$ in the
$i$-th row of the $t+1$-th column. \  $\bheteroparams \equiv \{ \bsigmasq , \btausq \}$ is the pair of
ragged $N$ by $T$ matrices, with $\ssigmasqit$ and $\stausqit$ in the $i$-th row of the $t$-th column
of $\bsigmasq$ and $\btausq$, respectively.
}} \ Further details and
an algorithm for implementation are provided in the appendix.

Consider the problem of estimating $\svolit$, the volatility parameters for person $i$ in year $t$, if all other parameters $\bvolnit$ (and $\bthetaphi$) were known. \
The decision tree for estimation is shown in Figure \ref{fig:hierarchy} and described here, both with references to relevant equations in the appendix.

\begin{enumerate}\label{hierarchy list}
       \item[Level 1] $\svolit$ can remain unchanged from last year ($\svolit=\svolitm1$, eq: \ref{eq: prevchoice1}) or can change ($\svolit \neq \svolitm1$, eq: \ref{eq: prevchoice2}). \ If $\svolit$ changes;
       \item[Level 2] $\svolit$ can change to a value from the set of \emph{other values for that individual}  ($\svolit \in \bvolint$ and $\svolit \ne \svolitm1$, eq: \ref{eq: personchoice1}) or can take on a value new to the individual ($\svolit \notin \bvolint$, eq: \ref{eq: personchoice2}). \ If $\svolit$ takes on a value new to the individual;
       \item[Level 3] $\svolit$ can be a value held by \emph{other individuals} ($\svolit\in\bvolnit$ and $\svolit \notin \bvolint$, eq: \ref{eq: popchoice1}) or can be a new value not shared with other individuals ($\svolit\notin\bvolnit$, eq: \ref{eq: popchoice2}).
\end{enumerate}
The probability that $\svolit$ takes a given value
is a function of a) the likelihood of generating estimated shocks $(\somegait,\svarepsilonit)$
given $\svolit$ and b) the prior probability of $\svolit$.

The prior probability that the parameter remains unchanged in Level 1 ($\svolit=\svolitm1$) is proportional to $ Q_{i,t}$; the prior probability that the parameter changes is proportional to $\theta$. \ If the parameter changes in Level 1 ($\svolit \neq \svolitm1$), the prior probability that $\svolit$ changes to a value held by that individual in another year in Level 2 is proportional to the number of times that value occurs in other years for that individual; the prior probability that $\svolit$ changes to a new value not seen for that individual in another year is proportional to $\Theta_{i}$. \ If the parameter changes to a new value not seen for that individual in another year in Level 2, the prior probability that $\svolit$ changes to one of the other population values in Level 3 is proportional to the number of times that value occurs within the population; the prior probability that $\svolit$ changes to a new value not seen elsewhere in the population is proportional to $\Theta$.

A detailed outline of this estimation algorithm is given in the appendix.  \
The appendix shows this compound prior algebraically, and also shows how it is combined
with the data to produce a posterior for $\svolit$. \
 We proceed iteratively through all $t$ within an individual and all $i$ across individuals. \
This entire scheme for choosing volatility values $\bheteroparams$ is nested within a larger
Gibbs sampling algorithm \citep{GemGem84}. \ This Markov Chain Monte Carlo (MCMC) approach simultaneously
estimates the other parameters of our model, namely shocks ($\bshockparams$) and income
coefficients ($\bhomoparams$).

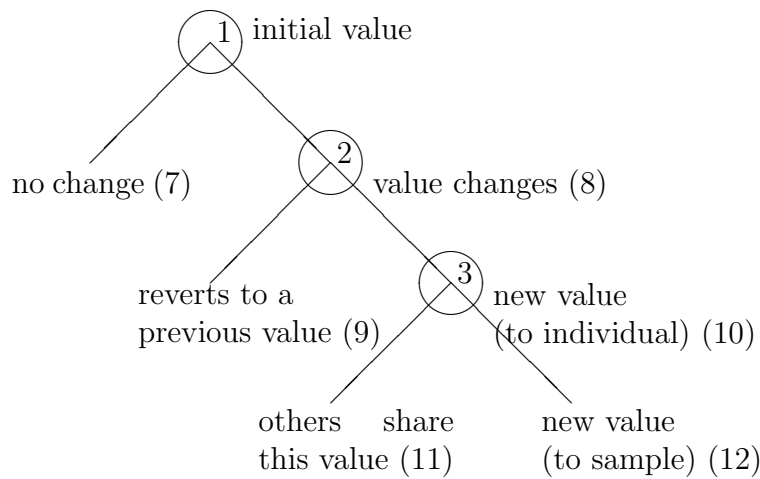
\begin{figure}[tp]
\caption{Model Hierarchy}
\label{fig:hierarchy}\centering
\setlength{\unitlength}{0.8cm}
\begin{picture}(6,7)
\put(1,6){\line(1,-1){2}}
\put(1,6){\line(-1,-1){2}}
\put(1,6){\circle{1}}
\put(1.1,6){1}
\put(1.7,6.1){\parbox{5cm}{initial value}}
\put(-2.3,3.5){\parbox{2.4cm}{no change (\ref{eq: prevchoice1})}}

\put(3,4){\line(1,-1){2}}
\put(3,4){\line(-1,-1){2}}
\put(3,4){\circle{1}}
\put(3.1,4){2}
\put(3.7,3.5){\parbox{5cm}{value changes (\ref{eq: prevchoice2})}}
\put(-0.2,1.3){\parbox{3.5cm}{reverts to a \\ previous value (\ref{eq: personchoice1})}}

\put(5,2){\line(1,-1){2}}
\put(5,2){\line(-1,-1){2}}
\put(5,2){\circle{1}}
\put(5.1,2){3}
\put(5.7,1.3){\parbox{3.5cm}{new value \\ (to individual) (\ref{eq: personchoice2})}}
\put(1.8,-0.8){\parbox{2.6cm}{others share this value (\ref{eq: popchoice1})}}
\put(6.5,-0.8){\parbox{3cm}{new value \\ (to sample) (\ref{eq: popchoice2})}}

\end{picture}
\put(-12,-3){\parbox{14cm}{Diagram describes evolution of volatility parameters. The numbers 1, 2, and 3 in circles
at each decision node correspond to the levels of the hierarchy described on page \pageref{hierarchy list}. \
The numbers (\ref{eq: prevchoice1}) through (\ref{eq: popchoice2}) identify the equation number
giving the probability of reaching that branch.}}
\put(-13,-4.5){\line(1,0){19}}
\end{figure}

\section{Results\label{section: results}}

{\singlespace
\begin{table}[t]
\caption[Basic Model Results]{Basic Model Results}
\label{table:basic results}
\begin{center}
{\normalsize
\begin{minipage}[t]{0.5\columnwidth}\centering
Distribution of Variance Parameters
\par
\begin{tabular}{lcc}
\hline\hline
& $%
\begin{array}{c}
\rm{Permanent} \\
\rm{Variance}%
\end{array}%
$ & $%
\begin{array}{c}
\rm{Transitory} \\
\rm{Variance}%
\end{array}%
$ \\ \hline

Mean                            & 0.0713  &       0.2771   \\ 
St. Dev.                        & 0.4685  &       1.0471   \\ 
N & 67,725 &       67,725 \\ \hline
1$^{\rm{st}}$ $\%$      & 0.0200  &       0.0499   \\ 
5$^{\rm{th}}$ $\%$      & 0.0250  &       0.0506   \\ 
10$^{\rm{th}}$ $\%$     & 0.0301  &       0.0510   \\ 
25$^{\rm{th}}$ $\%$     & 0.0313  &       0.0518   \\ 
50$^{\rm{th}}$ $\%$     & 0.0321  &       0.0530   \\ 
75$^{\rm{th}}$ $\%$     & 0.0331  &       0.0572   \\ 
90$^{\rm{th}}$ $\%$     & 0.0356  &       0.2452   \\ 
95$^{\rm{th}}$ $\%$     & 0.0498  &       1.2187   \\ 
99$^{\rm{th}}$ $\%$     & 0.8909  &       5.5030   \\ \hline
\end{tabular}
{\footnotesize Distribution of posterior means of $\bheteroparams$}
\end{minipage}\hfill
\begin{minipage}[t]{0.48\columnwidth}\centering
Shocks' Rate of Entry/Exit
\par
\begin{tabular}{ccc}
\hline\hline
lag & $\stheta_{\omega,k}$& $\stheta_{\varepsilon,k}$ \\ \hline
$k=0$   & 0.381        & 0.784        \\
  &  (0.088)    &  (0.029)     \\ 
$k=1$   & 0.865        & 0.180   \\
  &  (0.072)     &  (0.025)   \\ 
$k=2$   & 0.951        & 0.037      \\
  &  (0.064)     &  (0.017)\\ \hline
\end{tabular} \\
{\footnotesize
$\stheta_{\omega,k}$: impact of permanent shock \\ \ \ \ from $k$ periods ago \\
$\stheta_{\varepsilon,k}$: impact of transitory shock \\ \ \ \ from $k$ periods ago \\
Standard errors in parentheses. \\
}
\end{minipage}}
\end{center}
\par
{\footnotesize The left panel presents the posterior mean estimates of the
volatility parameters, $\bheteroparams$.
\ The distributions
presented here consider all years and all individuals together. \ The right
panel of this table present $\bthetaphi$, the mapping of shocks to income changes.  }
\end{table}
}

{\singlespace
\begin{figure}[t]
\caption[Distribution of Volatility Parameters]{Distribution of Permanent and Transitory Variance}
\label{fig:unconditionaldist}
\begin{center}
{\normalsize
\begin{tabular}{cc}
Permanent Variance & Transitory Variance \\
\includegraphics[width=2.8in]{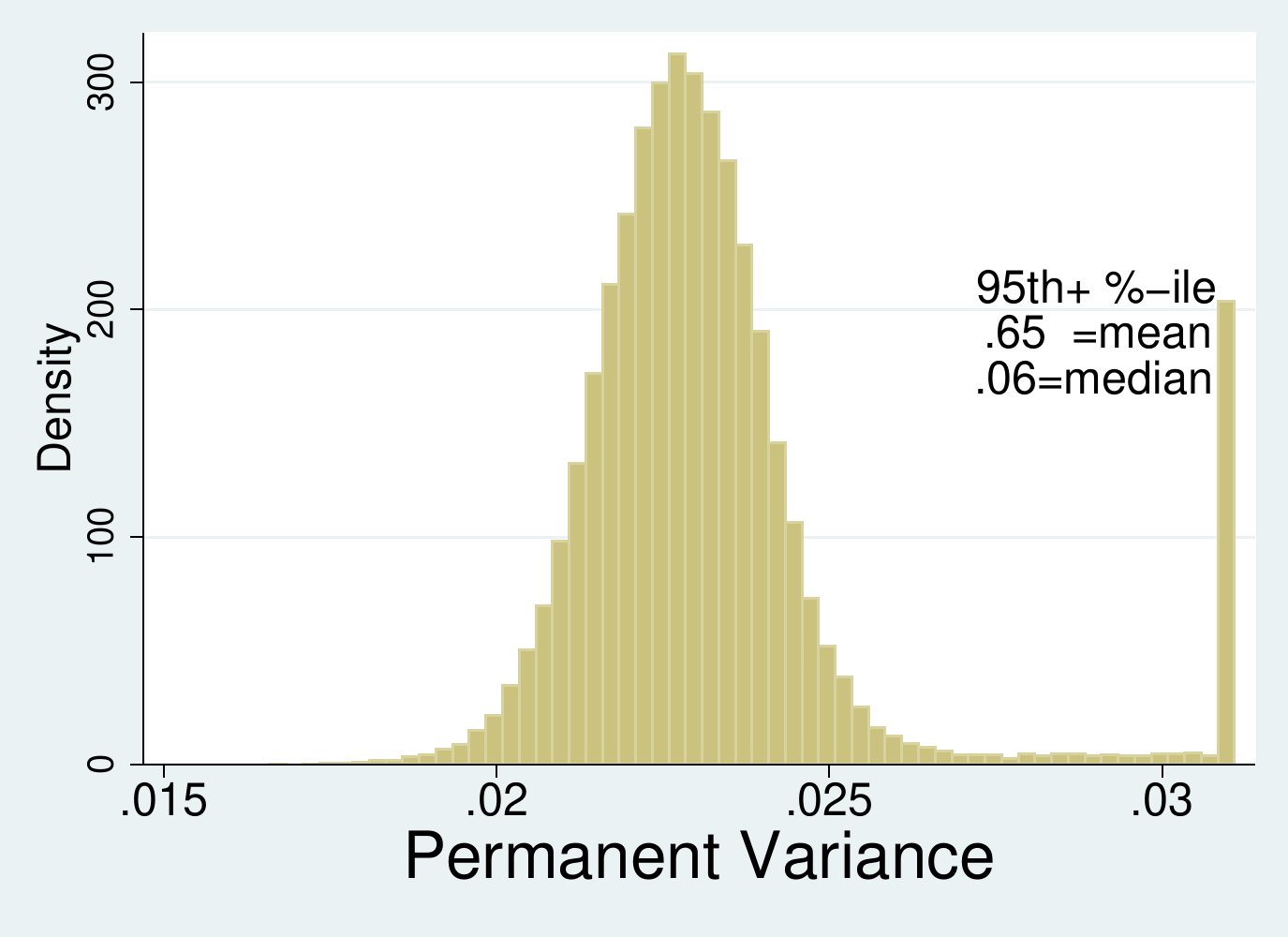} & %
\includegraphics[width=2.8in]{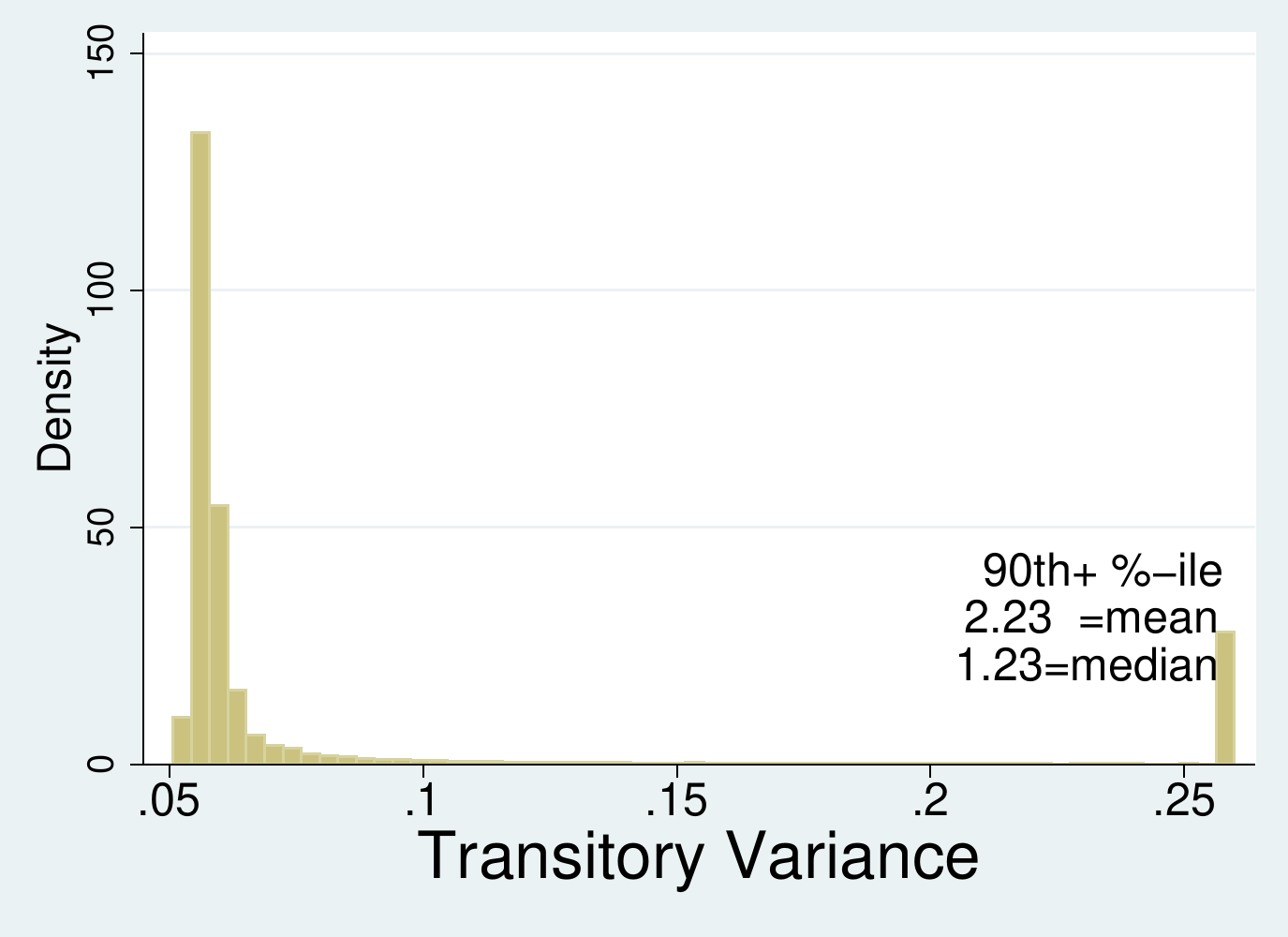}
\end{tabular}
}
\end{center}
\par
{\footnotesize This figure presents the distribution of $\btausq$ and
$\bsigmasq$. \ These are the distribution of posterior means
estimated from the data, as presented numerically in Table \ref{table:basic
results}. \ These posteriors of the permanent variance and transitory
variance are calculated for each individual in each year, as described in
Section \ref{section: estimation}. \ The distributions presented here
show all years and individuals together. \ Values are truncated at
the 95$^{\rm{th}}$ percentile for the permanent variance and at the 90th percentile for the transitory variance. \
Mean and median of the truncated part of each distribution is given. \\ \line(1,0){430}}
\end{figure}
}

Here, we present the model parameters estimated using the methods from
Section \ref{section: estimation}. \ The chief object of interest is
the evolution of the cross-sectional distribution of volatility
parameters, $\bvolt$, over time. \ These are shown in Section
\ref{subsection: pop evol results}. \ We begin with more basic results. \ In
subsection \ref{subsection: basic results}, we present estimates of the
homogeneous parameters $\bthetaphi$ that map shocks to income changes and
the unconditional distribution of volatility parameters, $\bheteroparams$. \ In Section
\ref{subsection: alternative explanations}, we rule out alternative explanations.
\ In Sections \ref{subsection: who is risky} and \ref{subsection: whose vol}, we
map these volatility parameter estimates to individuals' demographic or risk
attributes.

\subsection{Basic results\label{subsection: basic results}}

Table \ref{table:basic results} presents the basic parameter estimates
obtained from fitting our model to the PSID income data described in Section
\ref{section: estimation}. \ The left panel shows the distribution of risk
in the population, $\btausq$ and $\bsigmasq$. \ Formally, we
present the distribution of posterior means of permanent and transitory
variance parameters. \ The right panel show the mapping from shocks
to income changes, $\bthetaphi$, which we constrained to be constant
over time and across individuals.

{\singlespace
\begin{figure}[t]
\caption[Impulse Response Function]{Impulse Response Function for Permanent
and Transitory Shocks}
\label{fig:irf}
\begin{center}
{\normalsize
\begin{tabular}{cc}
Permanent Shock & Transitory Shock \\
\includegraphics[width=2.8in]{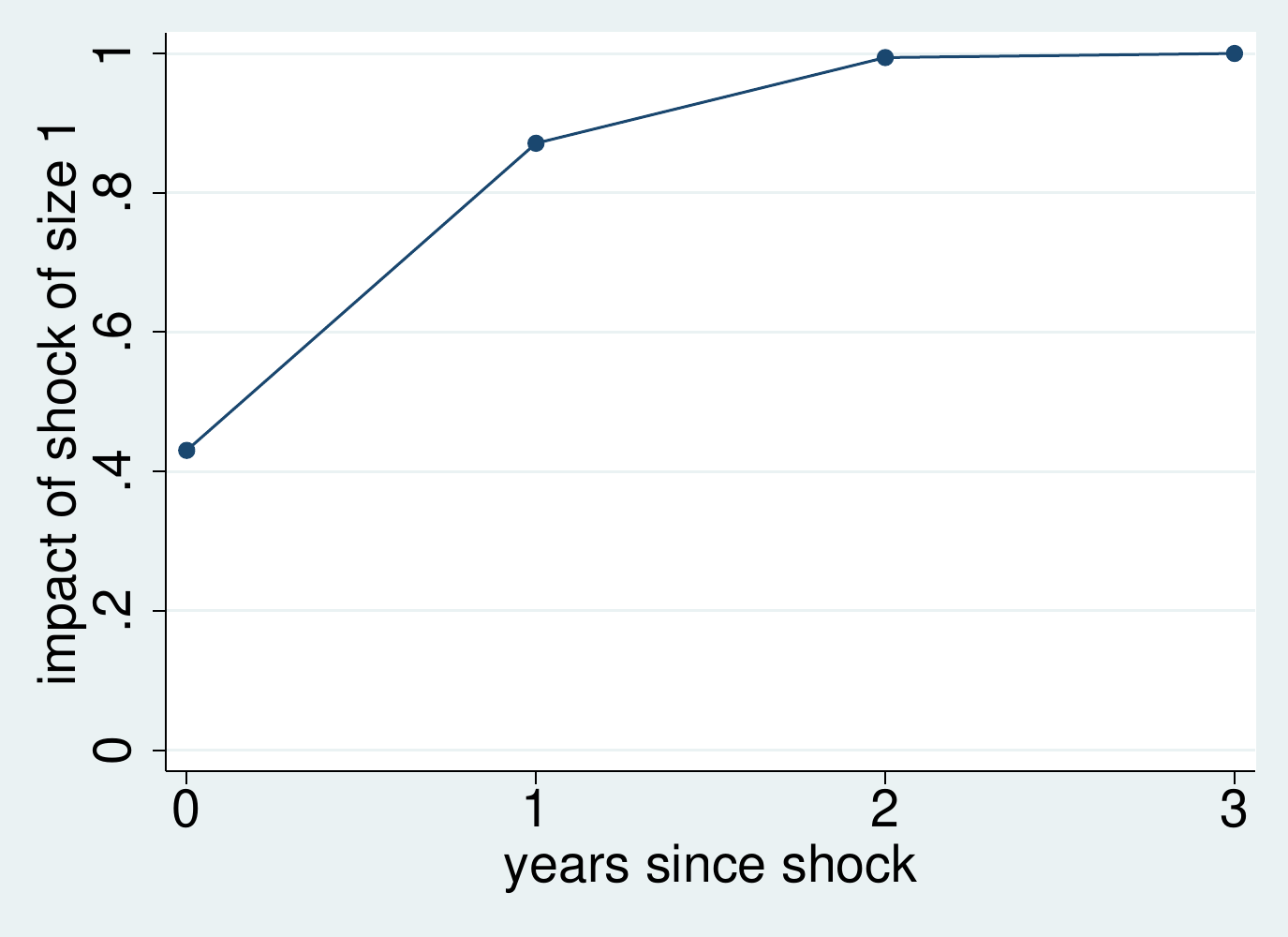} & %
\includegraphics[width=2.8in]{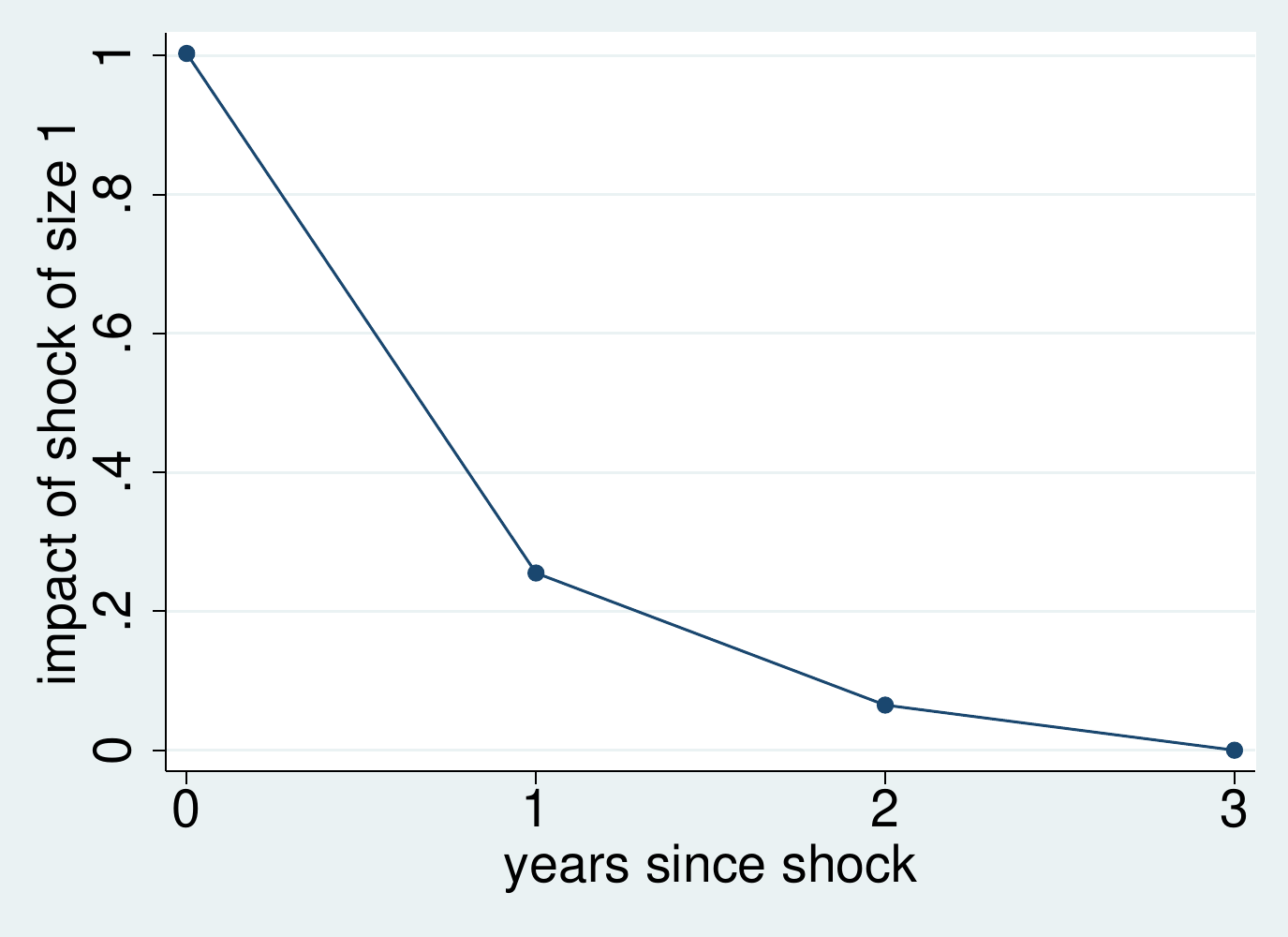}%
\end{tabular}
}
\end{center}
\par
{\footnotesize This figure presents an estimated impulse response function
for a permanent (left panel) and transitory (right panel) shock.  \line(1,0){430}}
\end{figure}
}

Note the extreme skew and fat tails (kurtosis) in the distribution of
volatility parameters, $\bheteroparams$, shown
in the left panel of Table \ref{table:basic results}). \ While medians are modest, means far exceed medians.
\ At the median, transitory shocks have a
standard deviation of approximately 23\% annually; permanent shocks have a
standard deviation of just under 18\% annually. \ However, the highest
volatility observations imply shocks with standard deviations well above
100\% annually. \ Figure \ref{fig:unconditionaldist} plots these skewed and
fat-tailed distributions by truncating the right tail.

As shown in the right panel of Table \ref{table:basic results},
permanent shocks enter in quickly ($\stheta _{\omega,k}$ are close to one) while
transitory shocks damp out quickly ($\stheta _{\varepsilon,k}$ fall to zero). \ The impact
of a shock on the evolution of income is presented in Figure \ref{fig:irf}.
\ These present impulse response functions for a permanent (left panel) and
transitory (right panel) shock. \ Shocks were calibrated as a one
standard-deviation shock for an individual with volatility parameters at the
estimated means (pulled from Table \ref{table:basic results}).

\subsection{Evolution of the volatility distribution\label{subsection: pop
evol results}}

{\singlespace
\begin{table}[p]
\caption[Year-by-Year Income Volatility Parameters]{Year-by-Year Income Volatility Parameters}
\label{table: volyby}\begin{center}
{\small
\begin{tabular}{ l c c c c c c c}
\hline\hline
 & \multicolumn{3}{c}{Permanent Variance, $\ssigmasq$} &&  \multicolumn{3}{c}{Transitory Variance, $\stausq$} \\ \cline{2-4} \cline{6-8}
  &       Mean &       Median &       95$^{\rm{th }}\%$ &&       Mean        &       Median &       95$^{\rm{th }}\%$ \\      
Average &   0.0713     &       0.0321  &       0.0498     &&     0.2771  &       0.0530  &       1.2186                \\      
\% Change    & 73$\%$    &   0$\%$  &       71$\%$          &&     99$\%$     &       1$\%$  &       154$\%$     \\      
Slope   &     0.0014             & 0.0000 &   0.0010 &&       0.0074  &       0.0000&       0.0508                  \\
(t-stat)        &  (6.84) &      (3.78)    &     (6.31)    &&      (7.02)    &     (9.37)    &    (6.25)                   \\      \hline  
1970    &      0.0573  &       0.0321  &       0.0424  &&       0.1568  &       0.0526  &       0.4498                  \\      
1971    &      0.0502  &       0.0321  &       0.0406  &&       0.1901  &       0.0526  &       0.6419                  \\      
1972    &      0.0411  &       0.0320  &       0.0374  &&       0.1909  &       0.0527  &       0.7775                  \\      
1973    &      0.0550  &       0.0321  &       0.0389  &&       0.2027  &       0.0528  &       0.7997                  \\      
1974    &      0.0481  &       0.0322  &       0.0437  &&       0.1848  &       0.0528  &       0.5520                  \\      
1975    &      0.0547  &       0.0321  &       0.0397  &&       0.1923  &       0.0530  &       0.7597                  \\      
1976    &      0.0663  &       0.0321  &       0.0464  &&       0.2746  &       0.0529  &       1.3527                  \\      
1977    &      0.0540  &       0.0321  &       0.0409  &&       0.2424  &       0.0529  &        1.1020                 \\      
1978    &      0.0557  &       0.0321  &       0.0411  &&       0.1865  &       0.0529  &       0.6785                \\      
1979    &      0.0738  &       0.0321  &       0.0432  &&       0.2226  &       0.0528  &       1.0134                  \\      
1980    &      0.0748  &       0.0321  &       0.0452  &&       0.2012  &       0.0529  &       0.7139                  \\      
1981    &      0.0651  &       0.0321  &       0.0504  &&       0.1986  &       0.0529  &       0.7762                  \\      
1982    &      0.0594  &       0.0321  &       0.0502  &&       0.2055  &       0.0529  &        0.8885                 \\      
1983    &      0.0744  &       0.0321  &       0.0457  &&       0.2550  &       0.0531  &       1.2691                 \\      
1984    &      0.0660  &       0.0321  &       0.0503  &&       0.2307  &       0.0531  &       0.9686                  \\      
1985    &      0.0593  &       0.0321  &       0.0477  &&       0.2260  &       0.0530  &        1.0063                 \\      
1986    &      0.0672  &       0.0321  &       0.0441  &&       0.2557  &       0.0529  &        1.1042                 \\      
1987    &      0.0679  &       0.0321  &       0.0477  &&       0.2448  &       0.0530  &         1.1468               \\      
1988    &      0.0714  &       0.0321  &       0.0467  &&       0.2286  &       0.0531  &        0.9494                \\      
1989    &      0.0629  &       0.0321  &       0.0490  &&       0.2462  &       0.0529  &        1.3182                 \\      
1990    &      0.0801  &       0.0321  &       0.0607  &&       0.2387  &       0.0530  &        0.9812                \\      
1991    &      0.0726  &       0.0321  &       0.0600  &&       0.2708  &       0.0530  &         1.2466              \\      
1992    &      0.0633  &       0.0321  &       0.0539  &&       0.2431  &       0.0531  &          1.0536               \\      
1993    &      0.0887  &       0.0321  &       0.0701  &&       0.4290  &       0.0532  &          2.6502               \\      
1994    &      0.0916  &       0.0321  &       0.0628  &&       0.4229  &       0.0532  &           2.3884              \\      
1995    &      0.0764  &       0.0321  &       0.0583  &&       0.4080  &       0.0532  &           2.2152              \\      
1996    &      0.0609  &       0.0321  &       0.0541  &&       0.4167  &       0.0531  &          2.4093               \\      
1997    &      0.0721  &       0.0321  &       0.0499  &&       0.3916  &       0.0531  &           2.3408              \\      
1999    &      0.0769  &       0.0321  &       0.0519  &&       0.3059  &       0.0532  &           1.5679              \\      
2001    &      0.0975  &       0.0322  &       0.0719  &&       0.2616  &       0.0531  &           1.0974              \\      
2003    &      0.1026  &       0.0322  &       0.0967  &&       0.4771  &       0.0534  &           2.4896              \\      
2005    &      0.1294  &       0.0324  &       0.0592  &&       0.4379  &       0.0538  &              2.2246           \\      \hline
\end{tabular}
}
\end{center}
\par
{{\footnotesize The construction of posterior means for $\ssigmasq$ and $\stausq$ for each individual in each year is detailed
in the text. \
The first row shows the full sample distribution, so that the second column shows the median value of the posterior mean
of $\ssigmasq$ over all individual-years. \ The second row shows the
percent change over the sample, calculated as the coefficient of
a weighted OLS regression of year-specific sample moments on a time trend, multiplied by the number of years (2005-1968) and
divided by the full sample value. \ The coefficient and t-statistic are shown below.  } }
\end{table}
}

Here, we show how the distribution of posterior means of variance parameters
has evolved over time. \ This evolution is shown in Tables \ref{table: volyby} and also
in Figure \ref{fig:voldistevol}. \ Table \ref{table: volyby} shows the year-by-year distribution
of volatility parameters ($\bvolt$)
posterior means. \ This table mirrors Table \ref{table: momentsyby},
with volatility parameter ($\svolit$)
posterior means replacing reduced form moments. \  The first three columns show results for the permanent variance parameter,
$\ssigmasq$; the final three columns show results for the transitory variance parameter, $\stausq$. \
The first and fourth columns present means of the permanent and transitory variance parameter posterior means,
the second and fifth
columns present medians of parameter posterior means, and the third and sixth columns present 95$^{\rm{th}}$ percentiles. \ All use weights from
the PSID. \ The first row shows whole-sample results. \ The second row shows the
percent change in the mean, median, or 95$^{\rm{th}}$ percentile over the sample.\footnote{\footnotesize{This is calculated
as coefficient of
a weighted OLS regression of the year-specific moments from below on a time trend, multiplied by the number of years (2005-1968)
and
divided by the whole-sample value in the previous row.}} \ The coefficient and t-statistic from this regression are shown just below. \
Year-by-year values are then shown.

Table \ref{table: volyby} shows that the mean of permanent and transitory parameters have increased substantially over the sample (by 73 and 99 percent,
respectively) while the
medians have not (0 and 1 percent increases, respectively). \ This divergence can be explained by an increase in the magnitude of
permanent and transitory variance parameters at the right tail,
among individuals with the highest parameters (the 95$^{\rm{th}}$ percentile values increasing 71 percent and 154 percent,
respectively). \ Colloquially, the kind of people whose incomes had always moved around a lot are
moving around even more than they used to; the median person's income does not move more than it used to. \

This pattern can be seen graphically in Figure \ref{fig:voldistevol}, which shows the year-by-year evolution of many quantiles
of the distribution of permanent and transitory variance posterior means. \
In the bottom panels of Figure \ref{fig:voldistevol}, we plot the 1st, 5$^{\rm{th}}$, 10th, 25$^{\rm{th}}$, 50th, and
75$^{\rm{th}}$ percentile values of the posterior mean of the permanent ($\ssigmasq$, left) and transitory ($\stausq$, right)
variance parameters by
year. \ These are very stable and show no clear upward trend. \ The size of
this increase is extremely small economically. \
Looking at all but the \textquotedblleft risky\textquotedblright tail of
the distributions, the distributions look very stable.

In the middle and upper panels of Figure \ref{fig:voldistevol}, we show the evolution of the \textquotedblleft
risky\textquotedblright\ tail of the distribution of posterior means. \ In
this case, variance parameters increase strongly and significantly. \ This
increase in the right tail of the distribution explains the increase in the
mean completely.

{\singlespace
\begin{figure}[p]
\caption[Evolution of Percentiles of Volatility Distribution]{%
Evolution of Percentiles of Volatility Distribution}
\label{fig:voldistevol}
\begin{center}
{\normalsize
\begin{tabular}{cc}
Permanent Income Changes & Transitory Income Changes \\
Mean and Median & Mean and Median \\
\includegraphics[height=1.65in]{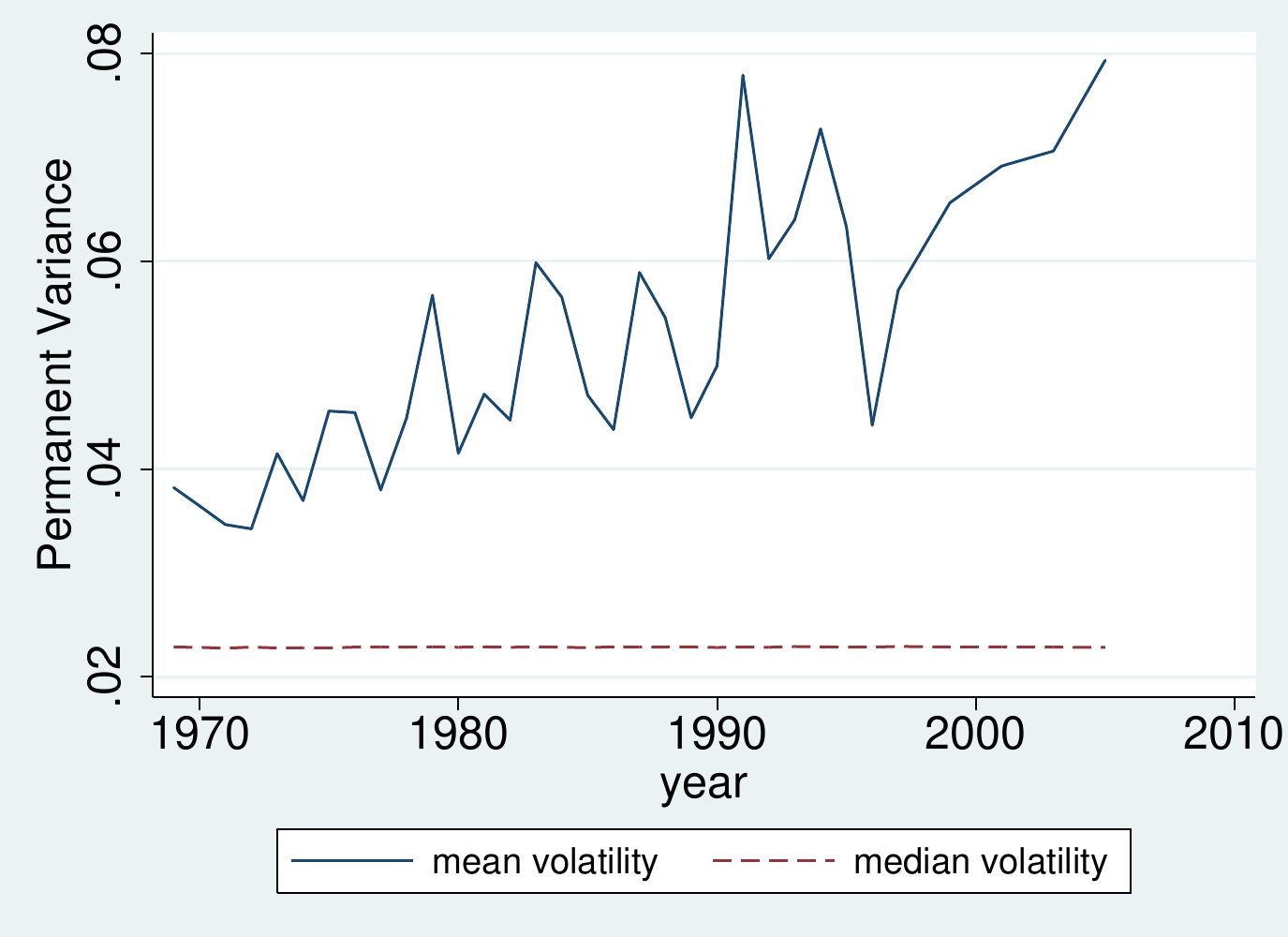} & %
\includegraphics[height=1.65in]{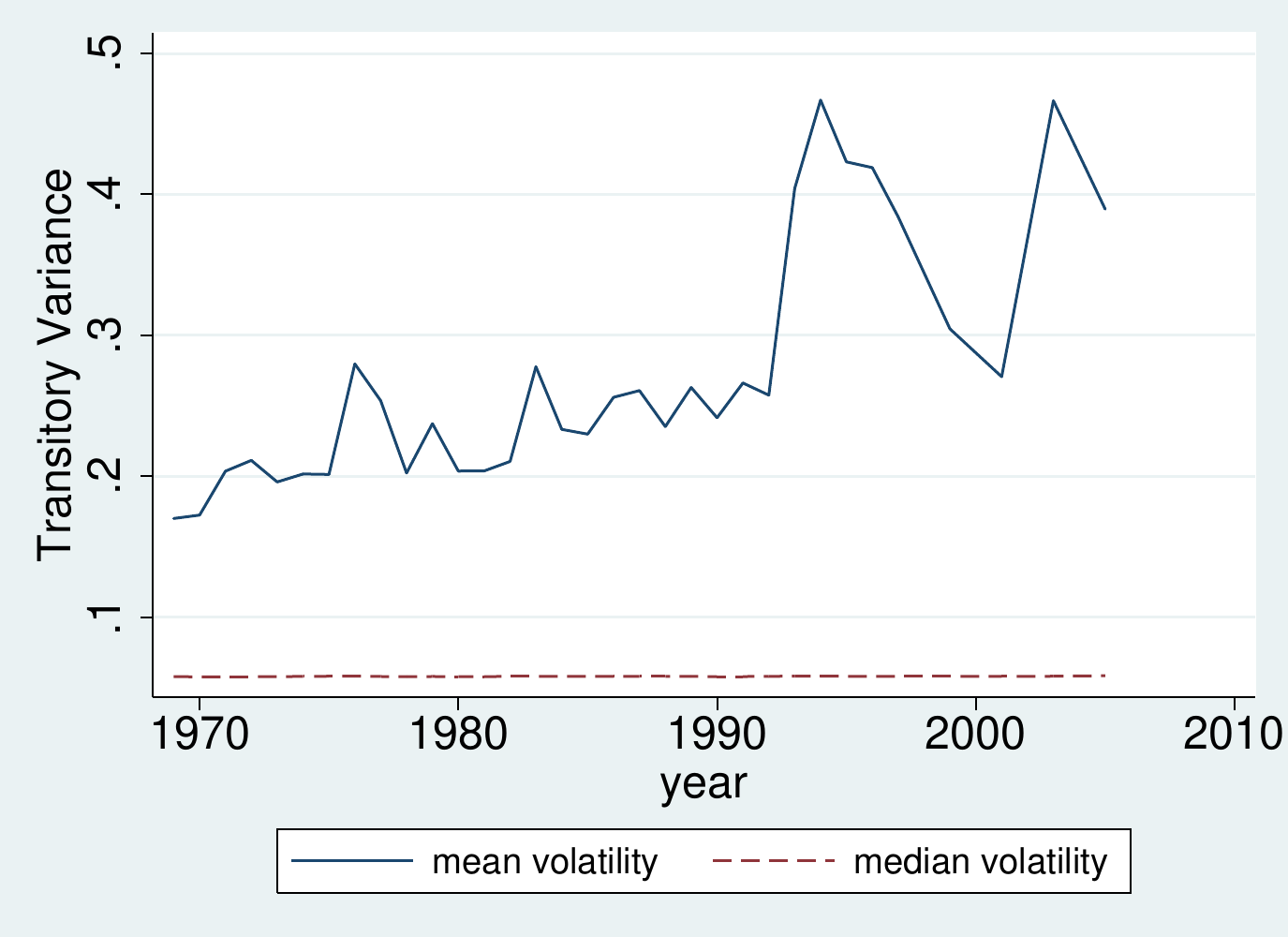} \\
$99^{\rm{th}}$ Percentile &  $99^{\rm{th}}$ Percentile \\
\includegraphics[height=1.65in]{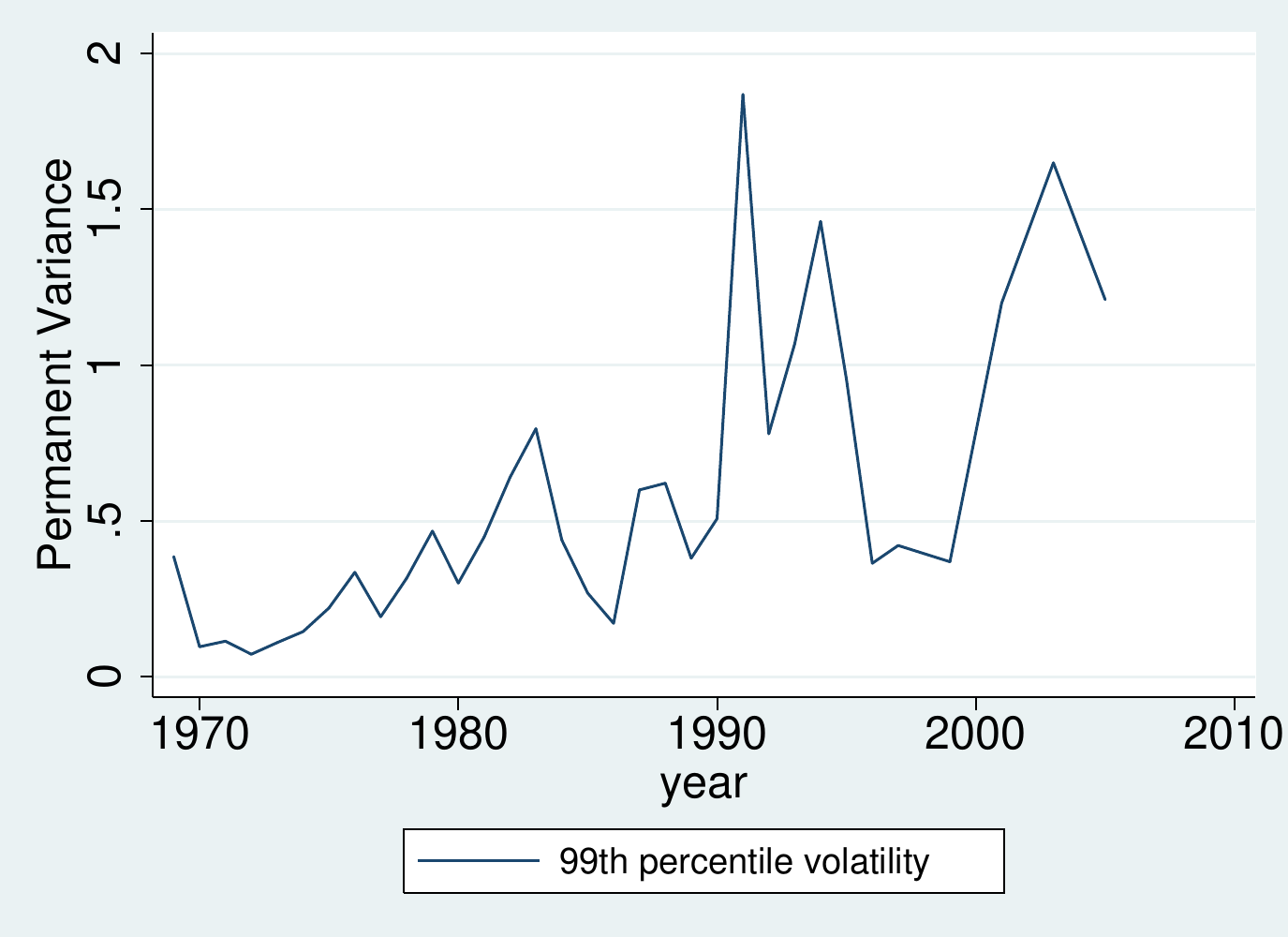} & %
\includegraphics[height=1.65in]{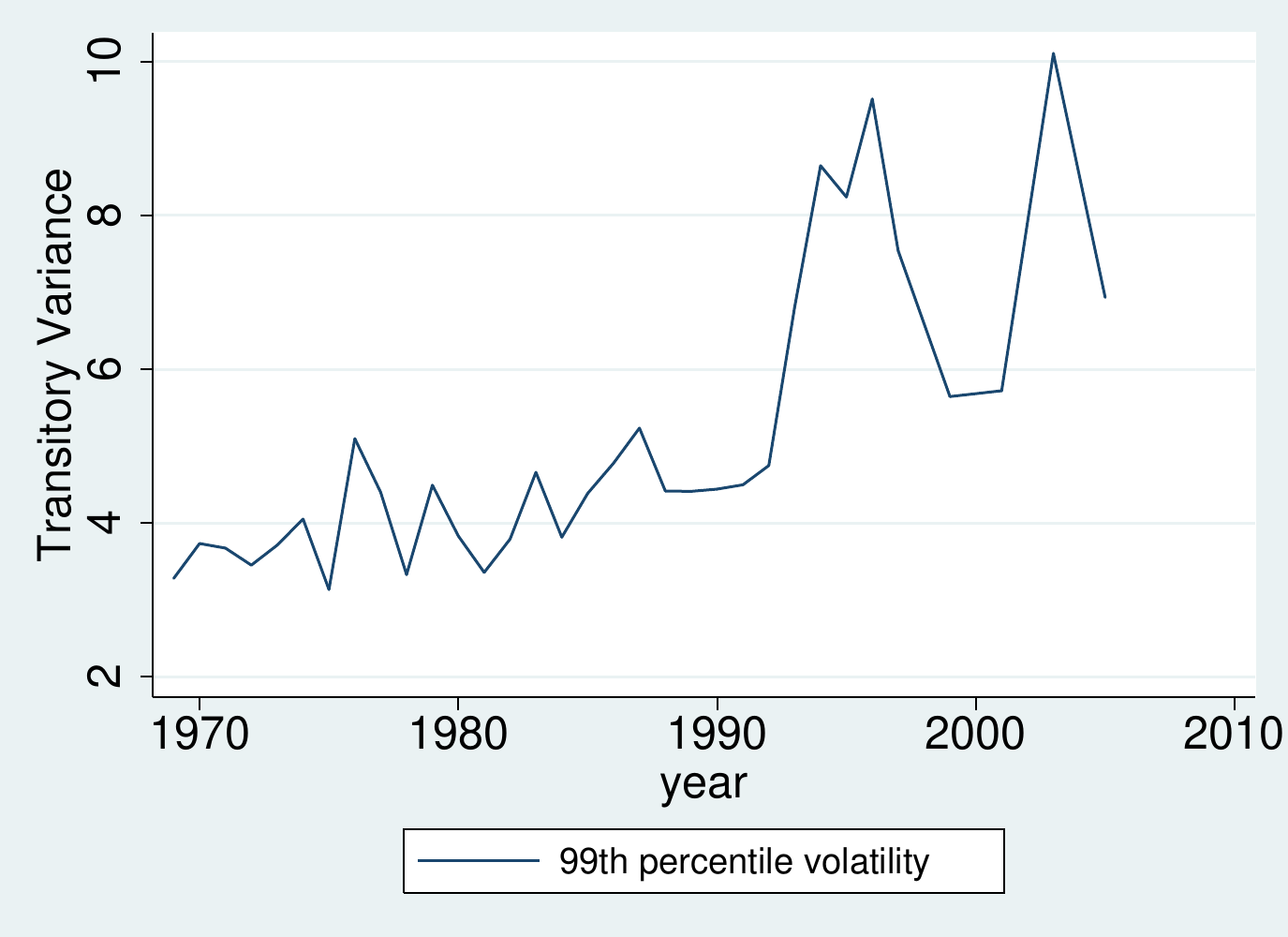} \\
$90^{\rm{th}}$ and $95^{\rm{th}}$ Percentiles & $90^{\rm{th}}$ and 95$^{\rm{th}}$ Percentiles \\
\includegraphics[height=1.65in]{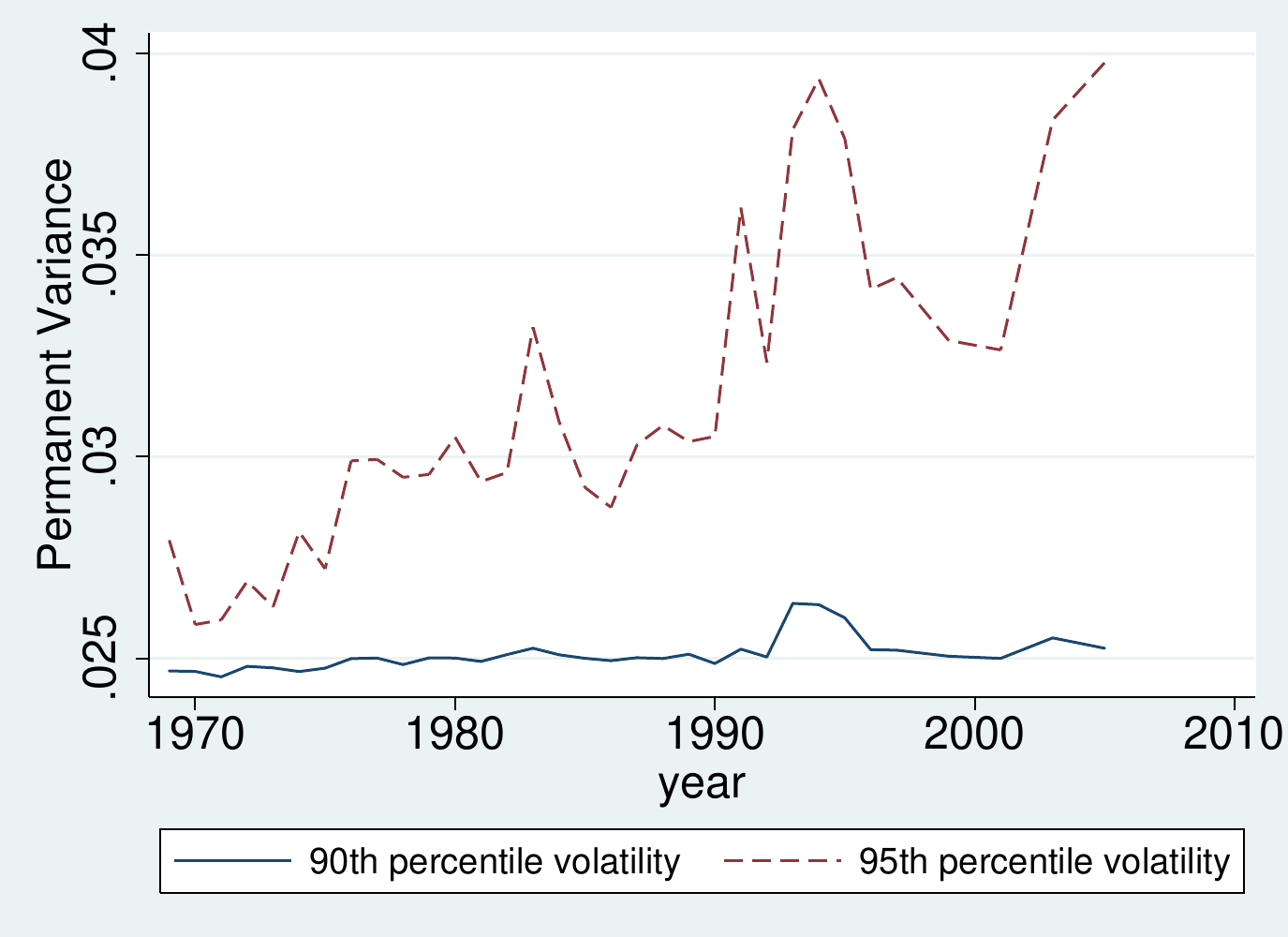} & %
\includegraphics[height=1.65in]{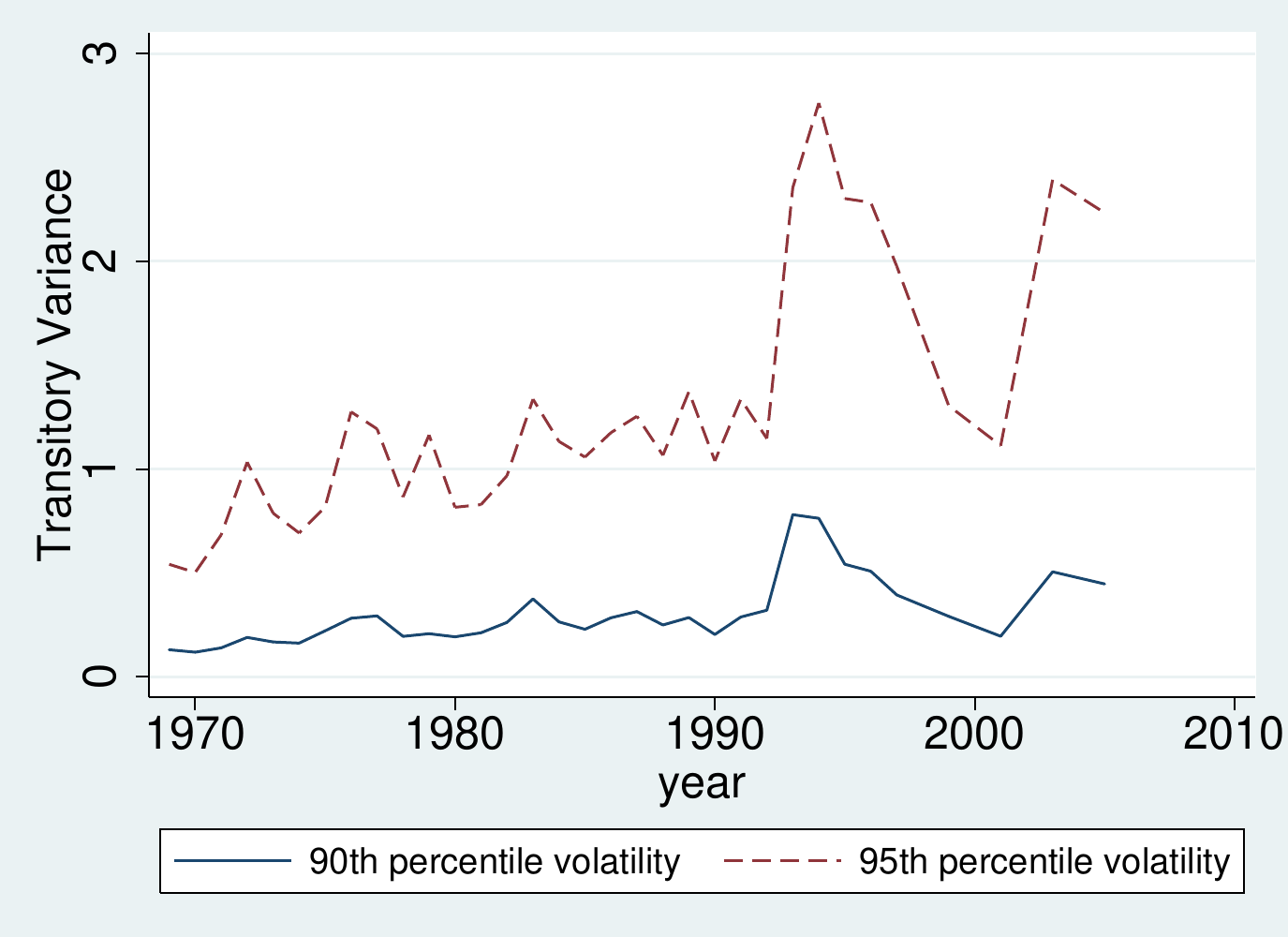} \\
$\le 75^{\rm{th}}$ Percentiles  &  $\le 75^{\rm{th}}$ Percentiles \\
\includegraphics[height=1.65in]{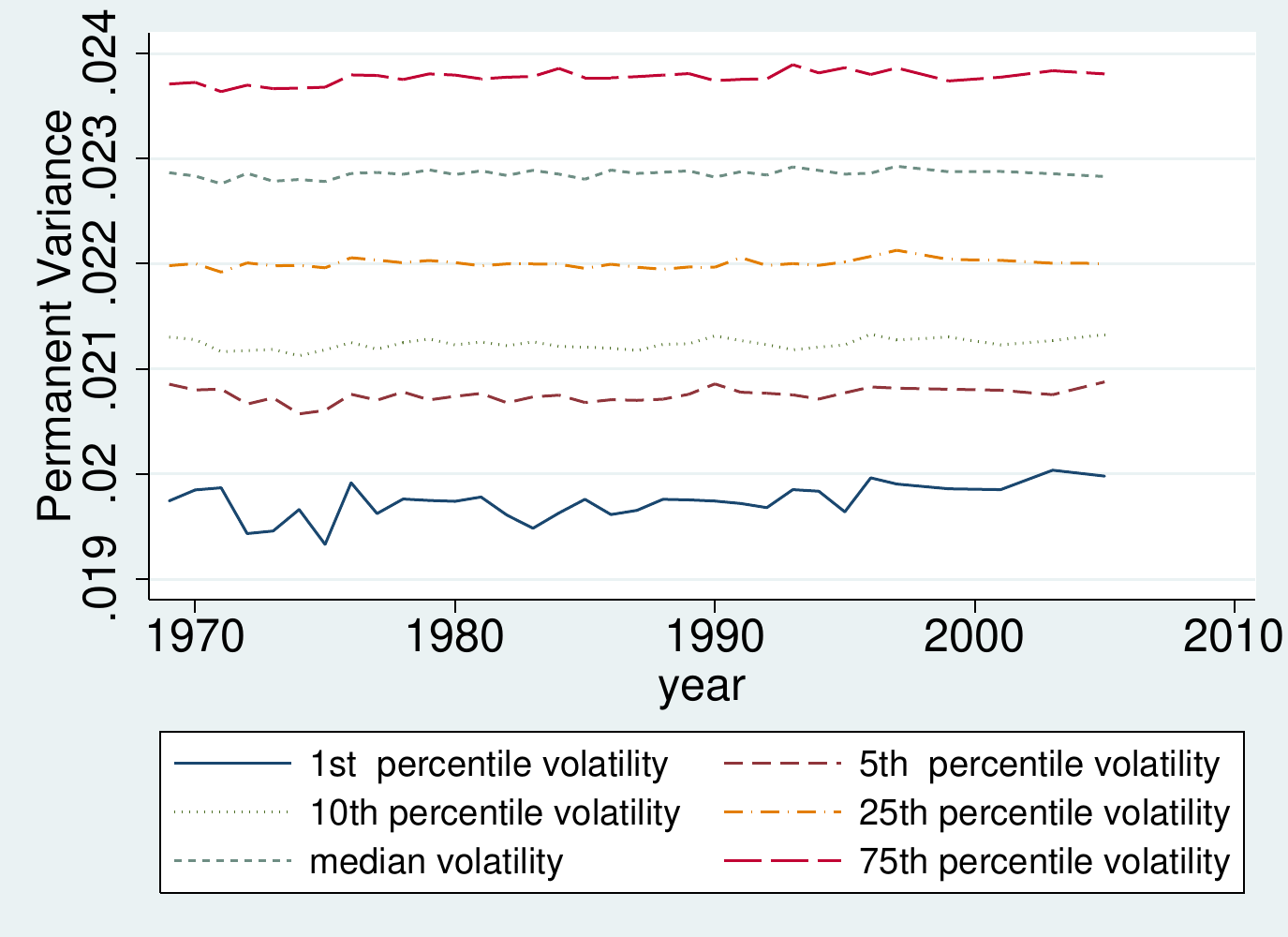} & %
\includegraphics[height=1.65in]{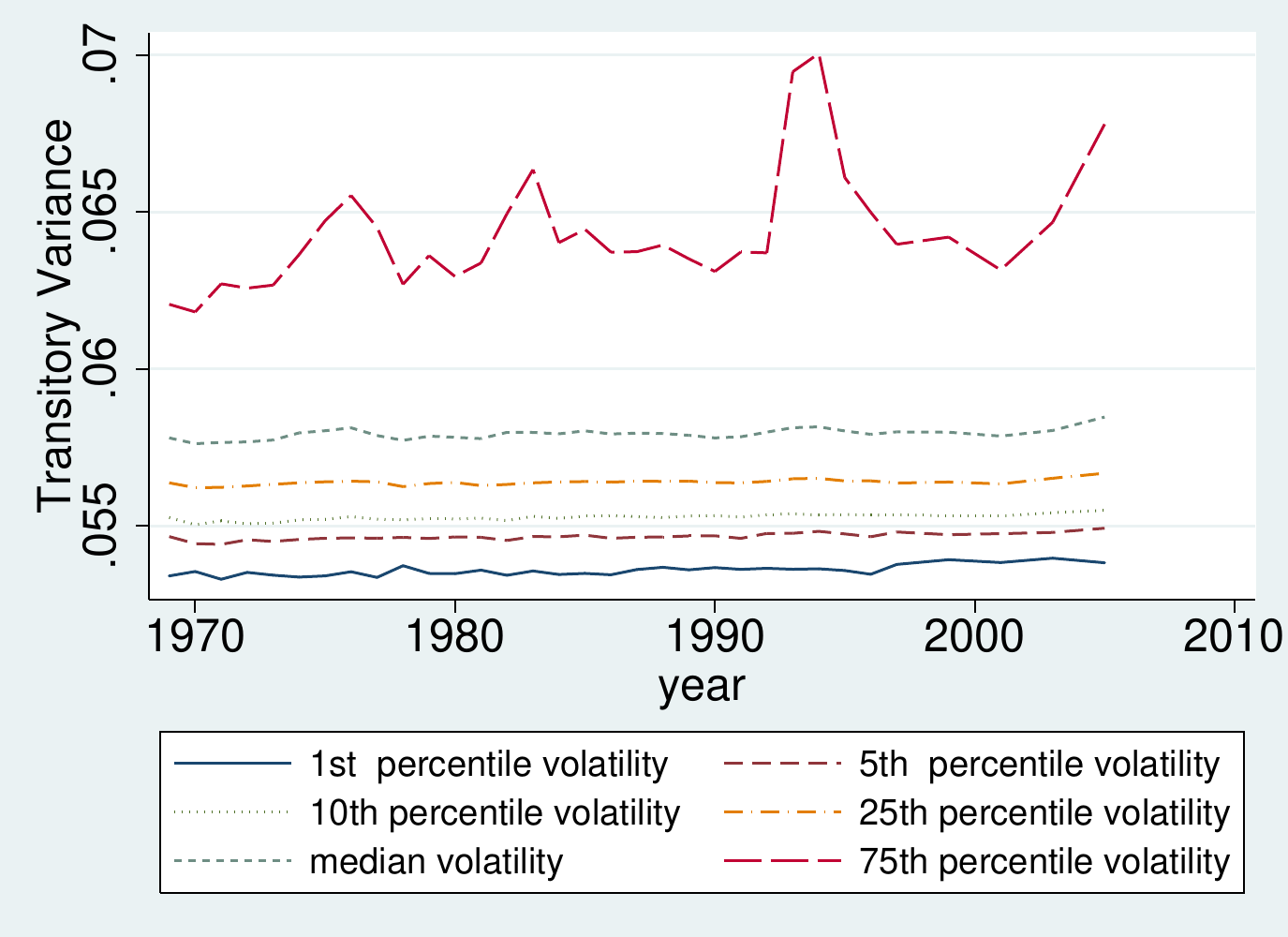}
\end{tabular}
}
\end{center}
\par
{\footnotesize These figures show the evolution of various percentiles of
the posterior mean of the permanent (left) and transitory (right) variance
for various percentiles of the distribution of variance parameters.
\\ \line(1,0){430}}
\end{figure}
}

\subsection{Heterogeneity or fat tails?\label{subsection: alternative explanations}}

So far, we have shown that the increases in income volatility
can be attributed solely to increases in the right tail of the volatility
distribution. \ To obtain this result, our model
assumes that the distribution of shocks is normal conditional on the volatility parameters. \
When the unconditional
distribution of shocks is fat-tailed (has high kurtosis), this is automatically attributed to
heterogeneity in volatility parameters. \ An alternative hypothesis is that there is little or no
heterogeneity in volatility parameters, but that shocks are conditionally fat-tailed.

When looking at the cross-section of income changes, heterogeneity in volatility parameters
(with conditionally normal shocks) and conditionally
fat-tailed shocks (without no heterogeneity in volatility parameters) are observationally equivalent;
they both imply a fat-tailed unconditional distribution
of income changes. \ By examining serial dependence, it is possible to reject the hypothesis
that everyone has the same volatility parameter. \ If shocks are conditionally fat-tailed but everyone has the same
volatility parameters, then those with large past income changes should be no more likely than others
to experience large subsequent income changes. \ If individuals differ in their volatility
parameters and those volatilities are persistent, then individuals with large past income changes
will be more likely than others to have large subsequent income changes.

This possibility is investigated in Table \ref{table: persistmomyby} and shown graphically in Figure \ref{fig:volpersist}. \
These compare the sample
variance of income changes for individuals with and without large past income changes. \
In each year, a cohort without large income changes is formed as the set of individuals whose measure of variance,
either permanent variance or squared income
change, was below median four years ago; a cohort with large income changes is formed as the set of individuals whose
measure of variance was above the 95$^{\rm{th}}$ percentile four years ago. \ This four-year period is chosen
so that income shocks are far enough apart to be uncorrelated. \citep{AbowdCard89}

Note that individuals with large past income changes tend to have larger subsequent income changes. \
The tendency to have large income changes is persistent, which indicates that some
individuals have \emph{ex-ante} more volatile incomes than others.

The divergence over time in volatility between past low- and high-volatility cohorts is clear in both
Figure \ref{fig:volpersist}
and Table \ref{table: persistmomyby}. \
The magnitude of income changes has been increasing more for those
with large past income changes (who are more likely to be inherently high-volatility) than for those
without such large past income changes (who are not).
\ This increase in volatility falls primarily on those who could be expected
to have volatile incomes to begin with. \ This shows that the increase in volatility
among the volatile we find in the model cannot be attributed to increasingly fat-tailed shocks
for everyone.

{\singlespace
\begin{table}[tbp]
\caption[Determinants of High Income Volatility]{Determinants of High Income Volatility (Probit)}
\label{table: probitreg}
\begin{center}
\begin{tabular}{lcc}
\hline\hline
Dependent & Permanent & Transitory \\
Variable & Variance & Variance \\ \hline
self-employed? 1 or 0 & 0.6001 & 0.7794 \\   & (24.07)*** & (32.22)*** \\   &  $[0.1085]$ &  $[0.1533]$ \\ 
risk-tolerant? 1 or 0 & 0.1303 & 0.0950 \\   & (5.91)*** & (4.31)*** \\   &  $[0.0180]$ &  $[0.0131]$ \\ 
age & 0.0104 & 0.0082 \\   & (7.82)*** & (6.20)*** \\   & $[0.0014]$ & $[0.0011]$  \\ 
years of education & -0.0041 & -0.0123 \\   & (-0.89) & (-2.67)*** \\   &  $[-0.0006]$ &  $[-0.0017]$ \\ 
income$>$median? 1 or 0& -0.2277 & -0.2922 \\   & (-9.84)*** & (-12.65)*** \\    & $[-0.0308]$  & $[-0.0398]$ \\ 
have children? 1 or 0 & -0.0498 & -0.0686 \\   & (-1.48) & (-2.04)** \\   &  $[-0.0068]$ &  $[-0.0094]$ \\ 
number of children & 0.0120 & 0.0068 \\   & (0.90) & (0.51) \\   &  $[0.0016]$ &  $[0.0009]$ \\ 
married? 1 or 0 & -0.1009 & -0.1815 \\   & (-3.00)*** & (-5.56)*** \\   & $[-0.0143]$  & $[-0.0270]$  \\ \hline
$R^{2}$ &  0.0469    & 0.0751 \\ 
observations & 31,898 & 31,898 \\ \hline
\end{tabular}
\end{center}

{\footnotesize Results from a probit regression to predict an indicator
variable for whether
posterior mean variance (permanent or transitory volatility) estimate is is above the 90$^{\rm{th}}$ percentile for that year. \
``Risk tolerant'' is set to 1 if the PSID risk tolerance variable exceeds 0.3. \  Above-median income indicates that
four-year lagged income is above-median for that (lagged) year.  \  *, **, and *** indicate significance at the 10$\%$, 5$\%$,
and 1$\%$ levels, respectively. \ z-statistics are in parentheses. \ Marginal effects are in square brackets.}
\end{table}
}

\subsection{Whose incomes are volatile?\label{subsection: who is risky}}

In this paper, we have identified increasing volatility for men in the U.S. since 1968 as
being driven solely by the right (volatile) tail of the volatility distribution. \
Here, we examine the attributes of men with highly
volatile incomes.

Table \ref{table: probitreg} presents the results from a probit regression to predict whether a person-year estimate of the (posterior mean)
volatility parameter is above the 90$^{\rm{th}}$ percentile for that year. \ Note from the first row
that self-employed individuals are
much more likely to have highly volatile incomes. \ The second row shows that ``risk tolerant'' individuals are
also much more likely to have highly volatile incomes. \ Risk tolerance is identified from answers to hypothetical
questions about lotteries, designed to elicit the individual's coefficient of relative risk-aversion; risk-tolerant individuals
are defined as those with an estimated coefficient of relative risk-aversion below 1/0.3. \

High income individuals (those with incomes above median four years before the observation in question)
are less likely to have volatile incomes. \ Individuals with more years of education are also less likely
to have volatile incomes. \ Older individuals are more likely to have volatile incomes, a result
driven by the large number of high-volatility individuals between ages 50 and 60. \
Unsurprisingly, men who are married and/or who have children are less likely to have volatile
incomes.

\subsection{Whose incomes are increasingly volatile?\label{subsection: whose vol}}

Section \ref{subsection: who is risky} identified attributes of individuals with volatile incomes. \
In particular, the self-employed and those whose answers to survey questions suggest they are risk-tolerant
are more likely to have volatile incomes. \ Here, we examine the increase in volatility over time among
these groups.

{\singlespace
\begin{table}[t]
\caption[Volatility Trends by Self-Employment, Income, and Risk Tolerance]{Volatility Trends by Self-Employment, Income, and Risk Tolerance}
\label{table: voltrendsample}
\begin{center}
{\footnotesize\setlength{\tabcolsep}{5pt}
Permanent Variance \par
\begin{tabular}{lcccccccc}
\hline\hline
& \multicolumn{2}{c}{Self-Employment}&&\multicolumn{2}{c}{Income}&&\multicolumn{2}{c}{Risk Tolerance} \\ \cline{2-3}  \cline{5-6}  \cline{8-9}
&           self-   &       not self-      &&  $>$ med. & $\le$ med.  &&   risk       &         not risk                   \\
sample        &       employed       & employed  &&  income &income    &&       tolerant     &       tolerant          \\  \hline
change per year&       0.0048  &       0.0011  && 0.0018 & 0.0009   &&   0.0035  &       0.0012  \\
$\%$ change '68-'05 &       194$\%$    &       58$\%$     && 135$\%$&36$\%$&&     172$\%$    &       76$\%$       \\
        &      \ (6.17)***       &      \ (4.58)***       && \ (5.99)***& \ (2.75)***&&      (4.61)***       &        (4.50)***         \\ \hline
N & 6,068 &   41,766  && 10,336 &23,876&&23,958 & 18,029  \\ \hline
\end{tabular}
}
\end{center}
\begin{center}
{\footnotesize\setlength{\tabcolsep}{5pt}
Transitory Variance \par
\begin{tabular}{lcccccccc}
\hline\hline
& \multicolumn{2}{c}{Self-Employment}&&\multicolumn{2}{c}{Income}&&\multicolumn{2}{c}{Risk Tolerance} \\ \cline{2-3}  \cline{5-6}  \cline{8-9}
&           self-   &       not self-      &&  $>$ med. & $\le$ med.  &&   risk       &         not risk                   \\
sample        &       employed       & employed  &&  income &income    &&       tolerant     &       tolerant          \\  \hline
change per year&       0.0262  &       0.0061  &&  0.0040 & 0.0116  &&   0.0100  &       0.0076  \\
$\%$ change '68-'05 &       176$\%$    &       101$\%$    &&   125$\%$ & 101$\%$&&    117$\%$    &       114$\%$       \\
 &       (11.27)***       &       (13.80)***       &&   (10.84)*** & (13.22)***&&    (7.81)***       &        (9.45)***         \\ \hline
N & 6,068 &   41,766  && 23,876 & 23,958&& 10,336  & 18,029  \\ \hline
\end{tabular}
}
\end{center}
\par
{\footnotesize Results from a weighted OLS regression to predict the
posterior mean variance (volatility) estimate with a linear time trend. \
The ``change'' row shows the coefficient on calendar time; the ``percent change'' row shows the
expected percent change over the sample implied by this coefficient. \ This is
(100 percent) times (2005 minus 1968) times (the coefficient on calendar time) divided by
(the average posterior mean in the sample). \
The top panel presents
results for the permanent variance; the bottom panel presents results for the transitory variance. \
Each column presents results for a different sub-sample. \ ``Risk tolerant'' means that
 the PSID risk tolerance variable exceeds 0.3. \  Above-median income indicates that
four-year lagged income is above-median for that (lagged) year.  \ t-statistics are in parentheses.  }
\end{table}
}

{\singlespace
\begin{table}[tbp]
\caption[Volatility Trends by Age and Education]{Volatility Trends by Age and Education}
\label{table: voltrendageedy}
\begin{center}
{\small
Permanent Variance
\begin{tabular}{lcccccc}
\hline\hline
 & \multicolumn{2}{c}{age}&& \multicolumn{3}{c}{education} \\ \cline{2-3} \cline{5-7}
        &       less than    &       at least       &&       more than       &      high   &       less than       \\
sample        &      40 yrs old   &      40 yrs old        &&high school  & school     &       high school     \\  \hline
mean change/year &       0.0006  &       0.0018  &&       0.0024  &       0.0005  &       0.0004  \\
$\%$ change '68-'05 &       44$\%$    &       76$\%$     &&       120$\%$     &       28$\%$     &       22$\%$     \\
        &       (3.66)***        &       (4.55)***        &&       (6.08)***        &        (1.71)* &       (1.17)        \\ 
median change/year&       0.0000  &       0.0000  &&       0.0000  &       0.0000  &       0.0000 \\
$\%$ change '68-'05 &       0$\%$    &       0$\%$     &&       1$\%$     &       0$\%$     &       0$\%$     \\
        &       (0.79)        &       (4.29)***        &&       (5.20)***        &        (-1.36) &       (0.20)        \\ 
95$^{\rm{th}}$ $\%$tile chnge/year &       0.0007  &       0.0010  &&       0.0008  &       0.0007  &       0.0012  \\
$\%$ change '68-'05 &       53$\%$    &       67$\%$     &&       63$\%$     &       55$\%$     &       72$\%$     \\
       &       (8.35)***        &       (6.47)***       & &       (10.32)***        &        (6.50)*** &       (2.31)**        \\ \hline
N & 23,928 &   23,906  && 23,455  & 15,516 &  8,863 \\ \hline

\end{tabular}
\newline
\setlength{\tabcolsep}{5pt}
\newline
\newline
Transitory Variance
\begin{tabular}{lcccccc}
\hline\hline
 & \multicolumn{2}{c}{age}&& \multicolumn{3}{c}{education} \\ \cline{2-3} \cline{5-7}
        &       less than    &       at least       &&       more than       &      high   &       less than       \\
sample        &      40 yrs old   &      40 yrs old        &&high school  & school     &       high school     \\  \hline
mean change/year &       0.0057  &       0.0096  &&       0.0093  &       0.0065  &       0.0066  \\
$\%$ change '68-'05 &       86$\%$    &       123$\%$     &&       120$\%$     &      102$\%$     &       95$\%$     \\
        &       (9.36)***        &       (13.27)***        &&       (12.14)***        &        (8.76)*** &       (6.91)***        \\ 
median change/year&      0.0000   &       0.0000  &&       0.0000  &       0.0000  &       0.0000  \\
$\%$ change '68-'05 &       1$\%$    &       2$\%$     &&       2$\%$     &       2$\%$     &       3$\%$     \\
        &       (6.87)***        &       (18.73)***        &&       (11.18)***        &        (13.69)*** &       (7.60)***        \\ 
95$^{\rm{th}}$ $\%$tile chnge/year &       0.0378 &       0.0649  &&       0.0598  &       0.0483  &       0.0467  \\
$\%$ change '68-'05 &       124$\%$    &       211$\%$     &&       183$\%$     &       188$\%$     &       135$\%$     \\
        &       (7.87)***        &       (17.10)***        &&       (12.15)***        &        (11.04)*** &       (5.78)***        \\ \hline
N &23,928 &   23,906 && 23,455  & 15,516 &  8,863 \\ \hline
\end{tabular}
}
\end{center}
{\footnotesize Results from a weighted OLS regression to predict the
posterior mean variance (volatility) estimate with a linear time trend. \
The ``change'' row shows the coefficient on calendar time; the ``percent change'' row shows the
expected percent change over the sample implied by this coefficient. \ This is
(100 percent) times (2005 minus 1968) times (the coefficient on calendar time) divided by
(the average posterior mean in the sample). \
The top panel presents
results for the permanent variance; the bottom panel presents results for the transitory variance. \
Each column presents results for a different sub-sample. \ t-statistics are in parentheses.  }
\end{table}
}

Table \ref{table: voltrendsample} predicts the posterior mean variance (volatility) estimates
described earlier with a linear time trend. \
The ``change'' row shows the coefficient on calendar time; the ``percent change'' row shows the
expected percent change over the sample implied by this coefficient. \ The top panel presents
results for the permanent variance; the bottom panel presents results for the transitory variance. \
Each column presents results for a different sub-sample. \ By comparing the first two columns,
note that that volatility has increased dramatically more for self-employed people than for others. \
These individuals have much higher average levels of volatility, but their percentage change in volatility
is still higher than for other individuals. \ Self-employed individuals account for
a substantial proportion of the overall increase in income volatility. \ Similarly,
the increase in permanent volatility (the variance of permanent shocks)
is much greater for those who self-identify as risk tolerant
(those whose estimated coefficient of relative risk aversion less than $1/0.3$) than those who do not. \
Transitory volatility does not show major differences in trend for risk tolerant and not risk tolerant individuals.

Table \ref{table: voltrendsample} shows that the increase in volatility is apparent throughout the
income distribution. \ While increases in the average variance of transitory shocks are
similar (in proportional terms) for
those with above- and below-median income, the variance of permanent shocks has increased more for
those with above-median income than for those with below-median income. \ While below-median
individuals are over-represented among those with the highest volatilities (Section \ref{subsection: whose vol}),
low income individuals are not driving the increase in volatility among those with the most volatile incomes.

Table \ref{table: voltrendageedy} presents results
by age and educational attainment. \ Note that while magnitudes vary, the increase in volatility
at the right
tail is present for those below and above 40, and across the education distribution. \

\section{Conclusion\label{section: conclusion}}

Increases in the size of income changes in the PSID can
be attributed almost entirely to the \textquotedblleft right tail\textquotedblright\
of the volatility distribution. \ Taking volatility as a proxy for risk,
those who would have had
risky incomes in the past now face even more risk.
\ Everyone else has had no substantial change.

Without knowing more, the welfare implications of this finding are unclear. \
Depending on what kind of people have volatile incomes, an increase
in volatility at the volatile end of the distribution could be more or less bad
than an increase in volatility for everyone. \
Consider the possibility (which we refute in Section \ref{subsection: who is risky}) that
risk tolerance is independent of income volatility or expected income. \
In this case, increasing volatility at the volatile end of the distribution decreases welfare more
than increasing risk throughout the distribution. \ When individuals have
decreasing absolute risk aversion, high levels of income risk (proxied here by volatility)
make people more vulnerable to additional risk. \citep{Gollier2001} \ If there is a compensating
differential for risk so that volatile incomes are also higher on average, then this effect
will be mitigated or reversed.

This paper shows that those with the most volatile incomes are also the most risk-tolerant. \
In this case, the increase in risk has hit those best able to handle it.
To the degree that income volatility is chosen (e.g., by choosing an occupation),
we would expect those with the highest tolerance for risk or the best risk-sharing opportunities
to take on the most volatile incomes. \ If it is these individuals whose volatility has
increased, it could blunt substantially any welfare costs associated with
increased income volatility.
\  Since the increase
in volatile has fallen disproportionately on the self-employed, it could also reflect an increase
in profitable (but volatile) business opportunities. \ In this case, there could even be welfare
gains associated with increased income volatility.

\singlespace
\pagebreak

\appendix\newpage

\section{Appendix A: Estimation}

We
estimate the joint posterior distribution of all unknown parameters
conditional on our observed data as:
\begin{equation}
p(\bhomoparams,\bshockparams,\bheteroparams|\by)\propto
p(\by|\bhomoparams,\bshockparams)\cdot
p(\bshockparams|\bheteroparams)\cdot p(\bheteroparams)  \label{eq: fullpost}
\end{equation}%
Following Bayes rule, the distribution of parameters given the data --
$p(\bhomoparams,\bshockparams,\bheteroparams|\by)$ -- is proportional to the product of the distribution
of the data given those parameters --
$p(\by|\bhomoparams,\bshockparams)$ -- and the probability of those parameters --
$p(\bshockparams|\bheteroparams)\cdot
p(\bheteroparams)$. \ We will estimate the posterior
distribution of our unknown parameters by Markov Chain Monte Carlo (MCMC)
simulation, specifically the Gibbs sampler. \citep{GemGem84} \ The Gibbs
sampler estimates the full posterior distribution in equation (\ref{eq:
fullpost}) by iteratively sampling a value for each unknown parameter
conditional on the current values of the other unknown parameters. \ In
other words, we iterate over the following steps.

\begin{enumerate}
\item[Step 1:] Sample new values of $(\bhomoparams)$ from
$p(\bhomoparams|\by,\bshockparams,\bheteroparams) $

\item[Step 2:]  Sample new values of $(\bshockparams)$, the shock parameters for each person and year, from
$p(\bshockparams|\by,\bhomoparams,\bheteroparams)$

\item[Step 3:]  Sample new values of $(\bheteroparams)$, the volatility parameters for each person and year, from
$p(\bheteroparams|\by,\bhomoparams,\bshockparams)$.
\end{enumerate}

These sampling steps form a Markov chain that is iterated until the set of
all parameters has converged to their joint posterior distribution.  This algorithm is programmed in Python and run on a grid cluster of computers. \
One run of this model (with 10,000 iterations) takes several weeks,
though multiple runs can be done simultaneously. \  Each of the runs was started from a randomly sampled set of initial parameter values. \
These multiple runs were used to evaluate convergence of the algorithm to a reasonable
set of samples from the posterior distribution of all parameters. \   The first 5000 iterations of each chain was discarded as the pre-convergence burn-in period, and our inference was based upon the remaining sampled values.  \

\subsection{Step 1: Sampling income process parameters ($\protect\bhomoparams$)}

\label{homoestimation}

In this step, we take realized shocks ($\bomega,\bvarepsilon$) as well as excess log income data ($\by$) as given,
to estimate the rate at which shocks pass through to income ($\bthetaphi$).
\ Reorganizing equation (\ref{eq: income
process}) and setting limits of $\sOmega=\sepsilon =3$
\citep[a conservative choice according to][]{AbowdCard89}, we get
the following dynamic linear model,
\begin{equation}
y_{i,t}=
       \sum\limits_{k=0}^{t-3}\somega _{i,k}
       +\sum\limits_{k=t-2}^{t}\stheta _{\omega,t-k}\somega _{i,k}
       +\sum\limits_{k=t-2}^{t}\stheta _{\varepsilon,t-k}\svarepsilon _{i,k}
\label{eq: dynamiclinearmodel}
\end{equation}
For each individual $i$, the dynamic linear model for their excess log
income ($\by_{i}$) is a combination of the homogeneous
parameters $(\bthetaphi)$ and realized shocks
($\bshockparams$).
In our Gibbs sampling model
implementation, we take advantage of the fact that sampling new values of
the homogeneous parameters conditional on fixed values of the realized
shocks is relatively simple, and vice versa.

If we are given values of the realized shocks
($\bomega _{i},\bvarepsilon_{i}$),
we can calculate the scalar $y_{i,t}^{\star }$
and the $1\times 6$ (since $\sOmega+\sepsilon$=6) vector $X_{i,t}$ ,

\begin{equation*}
       y_{i,t}^{\star }\equiv
       y_{i,t}-\sum\limits_{k=0}^{t-3}\somega _{i,k}
       \qquad \qquad
       X_{i,t} \equiv
       (\somega _{i,t-2},\somega _{i,t-1},\somegait,
       \svarepsilon_{i,t-2},\svarepsilon _{i,t-1},\svarepsilonit)
\end{equation*}
Let $\by^{\star }$ be the $N(T-3)\times 1$ vector of all
$y_{i,t}^{\star }$ across individuals $i$ and time $t$, and let $\bX$
be the $N(T-3)\times 6$ matrix whose rows are all
$X_{i,t}$ across individuals $i$ and time $t$. We can then write
equation (\ref{eq: dynamiclinearmodel}) as a simple linear regression model,
\begin{equation*}
       \by^{\star }=
       \bX \cdot \bbeta + \be
       \qquad \mathrm{where} \quad
       \be \sim \mathrm{Normal}(\bzero,\sgammasq \cdot \bI)
\end{equation*}
where $\bbeta=(\stheta _{\omega,2},\stheta _{\omega,1},\stheta _{\omega,0},
\stheta _{\varepsilon,2},\stheta _{\varepsilon,1},\stheta _{\varepsilon,0})$
are the homogeneous parameters of interest. \ Note that this is the
stage at which we use measurement error (\be) as distinct from transitory shocks.

We use non-informative prior distributions for both $%
\sgammasq$ and $\bbeta$, which leads to the following
posterior distributions (the Bayesian analog of a least-squares estimate):
\begin{eqnarray}
\sgammasq &\sim &\mathrm{Inv-Gamma}\left( \frac{TN}{2}\,,\,\frac{(\by%
^{\star }-\bX\hat{\beta})^{\prime }(\by^{\star }-\bX\hat{%
\bbeta})}{2}\right)  \notag \\
\bbeta &\sim &\mathrm{Normal}\left( \hat{\bbeta}\,,\,\gamma
^{2}\cdot (\bX^{\prime }\bX)^{-1}\right)  \label{eq: regressionpost}
\end{eqnarray}%
where $\hat{\bbeta}=(\bX^{\prime }\bX)^{-1}\bX^{\prime }%
\by^{\star }$ as in a least-squares regression. \ We
sample new values of $\sgammasq$ and $\bthetaphi$ from the
distributions in (\ref{eq: regressionpost}), but with the additional
constraint that $\sum_{k}\stheta _{\varepsilon,k}=1$.

\subsection{Step 2: Sampling realized shocks ($\protect\bshockparams$)}\label{shockestimation}

In this step, we take excess log income data ($\by$),
the homogeneous parameters ($\bthetaphi$), and the
volatility parameters $(\bheteroparams)$ as given. \ We use these to sample realized shocks ($\bshockparams$).

If we are now given values of the homogeneous parameters
$(\bthetaphi)$, then the only unmeasured variables in our dynamic linear
model (\ref{eq: dynamiclinearmodel}) are the realized shocks
($\bomega_{i},\bvarepsilon _{i}$). \ We use maximum likelihood
estimates from a Kalman filter \citep{Kal60} to sample new values of
the realized shocks ($\bomega _{i},\bvarepsilon _{i}$), as
outlined in \citet{CarKoh94}. \ Given the homogeneous parameters
$(\bhomoparams)$ and the collection of volatility parameters
$(\bheteroparams)$, each individual's income process is independent,
so run the Kalman filter and sampling procedure for the
realized shocks ($\bomega _{i},\bvarepsilon _{i}$) for each
individual $i$ separately.

\subsection{Step 3: Sampling volatility parameters ($\protect\bheteroparams$)}

\label{heteroestimation}

In this step, we take sampled realized shocks ($\bshockparams$) as given and use these to sample
estimates of volatility parameters ($\bheteroparams$). \  In order to sample a full set of volatility parameters
$\bheteroparams$ from the distribution
$p(\bheteroparams|\by,\bhomoparams,\bshockparams)$, it is easiest to proceed sequentially by sampling (one-by-one),
the volatility parameters $\svolit$ for individual $i$ and year $t$ from the distribution
$p(\bheteroparams|\by,\bhomoparams,\bshockparams,\bvolnit)$. \ Note that under this scheme, information about ($\svolit$)
comes from our sampled permanent and transitory shocks
($\sshockit$) as well as our current estimates of the volatility parameters, $\bvolnit$, from other years within
the individual as well as other individuals. \  We link these other volatility values $\bvolnit$ to our shock
parameters $(\sshockit)$ through the posterior distribution,
\begin{equation}
p(\svolit|\sshockit,\bvolnit)
\propto p(\sshockit|\svolit)
\cdot p(\svolit|\bvolnit)
\label{eq: mhdpposterior}
\end{equation}%
The first term of equation (\ref{eq: mhdpposterior}) comes from the
likelihood of our realized shocks $(\sshockit)$ from
our dynamic linear model,
\begin{equation}
p(\sshockit|\svolit)\propto
\left( \ssigmasqit\stausqit\right) ^{-\frac{1}{2}}\exp \left(-
\frac{1}{2}\frac{\somegait^{2}}{\ssigmasqit}-\frac{1}{2}\frac{\svarepsilonit^{2}}{\stausqit}
\right)
\label{eq: mhdplikelihood}
\end{equation}%
The second term of equation (\ref{eq: mhdpposterior}) is our Markovian
hierarchical Dirichlet process (MHDP) prior,
$p(\svolit|\bvolnit)$, described in Sections \ref{section: heterogeneity} and \ref{section: estimation}.
Sampling new values $\svolit$ from the posterior
distribution (\ref{eq: mhdpposterior}) is a multi-step process that
acknowledges the structure of our population.
First, we sample a volatility parameter proposal
value ($\ssigsqstar\equiv\{\ssigmasqstar,\stausqstar\}$) from a continuous
distribution $f(\cdot )$.  For our implementation, we used an inverse-Gamma distribution, which is commonly used for
variance parameters. \ We will set
$\svolit=\ssigsqstar$
only if we cannot find a suitable $\svolit \in \bvolnit $ i.e. among our currently existing values in the population.

\subsubsection{Level 1: Is volatility unchanged from last year?}

We first consider the posterior probability that $\svolit=\svolitm1$,
\begin{eqnarray}
p(\svolit = \svolitm1) &
\propto & Q_{i,t} \cdot p( \sshockit | \svolitm1)  \label{eq: prevchoice1} \\
p(\svolit \ne \svolitm1) &
\propto & \theta \ \ \ \cdot p( \sshockit | \ssigsqstar)
\label{eq: prevchoice2}
\end{eqnarray}
Recall that $Q_{i,t}$ is the number of consecutive years
with parameter values $\svolitm1$; $\theta$ is the prior tuning parameter for Level 1. \
We compare the posterior probability that volatility values are unchanged from last year
($\svolitm1$ in equation (\ref{eq: prevchoice1}))
to the posterior probability that volatility values are equal to the
proposal value ($\ssigsqstar$ in equation (\ref{eq: prevchoice2})). \
We sample a possible value for $\svolit$ from this posterior distribution, either
$\svolitm1$ or $\ssigsqstar$, where
choice is made stochastically by flipping a weighted coin with weights
equal to the probabilities in equations (\ref{eq: prevchoice1}) and (\ref{eq:
prevchoice2}). \ If this weighted coin flip selects $\svolitm1$, then we set
$\svolit=\svolitm1$. \ If the coin flip selects $\ssigsqstar$, we do not set $\svolit=\svolitm1$ and instead proceed to Level 2 to find $\svolit$.

\subsubsection{Level 2: Is volatility the same as in another year?}

Given that we did not choose to set $\svolit \ne \svolitm1$, we consider the posterior probability that
$\svolit \in \bvolint$. \ If there are $\sNi$ unique values $\svolni \in \bvolint$, the posterior
probability that $\svolit=\svolni$ is,
\begin{eqnarray}
p(\svolit = \svolni) & \propto &
n_l \cdot p( \sshockit | \svolni) \quad l = 1, \ldots, \sNi  \label{eq: personchoice1} \\
p(\svolit \notin \,\,
\bvolint) & \propto & \Theta_{i} \cdot p( \sshockit | \ssigsqstar)  \label{eq: personchoice2}
\end{eqnarray}
$n_l$ is the number of occurrences of value $\svolni$ within the set of possible values $\bvolnit$;
$\Theta_{i}$ is the prior tuning parameter for Level 2.  \
We sample one of these $\sNi +1$ choices by flipping a weighted coin with weights proportional to the probabilities
above. \  If this weighted coin flip selects $\svolni \in \bvolint$, then we set
$\svolit=\svolni$.  \
If the coin flip selects $\ssigsqstar$, we do not set $\svolit=\svolni$ for any
$\svolit \notin \bvolint$ but instead proceed to Level 3 to find $\svolit$.

\subsubsection{Level 3: Is volatility the same as another person's?}

Given that $\svolit \notin \bvolint$, we consider the posterior probability that
$\svolit \in \pmb{\ssigma^{2}_{-i}}$, where $\pmb{\ssigma^{2}_{-i}}$ are the volatility
values that currently exist in the population outside of individual $i$.  \
If there are $\sN$ unique values $\svoln \in\pmb{\ssigma^{2}_{-i}}$, the posterior
probability that $\svolit=\svoln$ is,

\begin{eqnarray}
p(\svolit = \svoln) & \propto & n_l
\cdot p( \sshockit | \svoln) \qquad
l = 1, \ldots, \sN  \label{eq: popchoice1} \\
p(\svolit  \notin  \,\,
\bvolnit ) & \propto & \Theta \cdot p( \sshockit |
\ssigsqstar)  \label{eq: popchoice2}
\end{eqnarray}
$n_l$ is the number of occurrences of $\svoln$ within the set of current volatility values over all people other
than person $i$; $\Theta$ is the prior tuning parameter for Level 3.\
We sample one of these $\sN +1$ values by flipping a weighted coin with weights proportional to the probabilities
above. \  If this weighted coin flip selects $\svoln \in \bvolnit$, then we set
$\svolit=\svoln$.  \
If the coin flip selects $\ssigsqstar$, we set $\svolit=\ssigsqstar$. \
$\ssigsqstar$ represent new volatility values that have not yet been seen in the population.

\vspace{.1in}

The three steps outlined above result in a sampled volatility value $\svolit$ for person $i$ and year $t$,
conditional on the other volatility values $\bvolnit$.  \
We can repeat this procedure for all other years and individuals to update our full set of volatility values $\bheteroparams$.

\pagebreak
\bibliographystyle{econometrica}
\bibliography{shsrefs30}
\end{document}